\documentclass{jfm}
\usepackage{amsmath}
\usepackage[hidelinks]{hyperref}
\usepackage{graphicx}
\usepackage{epstopdf, epsfig}
\usepackage{xcolor}
\usepackage{url}

 \usepackage{comment}
 \shorttitle{Inertial soluble Marangoni flow}
 \shortauthor{Jun Eshima, Luc Deike, and Howard A. Stone}

\title{The effects of surfactant solubility on inertial Marangoni flow: theory and numerics}

\author{Jun Eshima\aff{1}\corresp{\email{jeshima@princeton.edu}}, Luc Deike\aff{1,2}\corresp{\email{ldeike@princeton.edu}} \and
    Howard A.  Stone\aff{1}\corresp{\email{hastone@princeton.edu}}}

\affiliation{$^1$ Department of Mechanical and Aerospace Engineering, Princeton University, Princeton, New Jersey 08544, USA \\
$^2$ High Meadows Environmental Institute, Princeton University, Princeton, New Jersey 08544, USA} 
\begin{document}

\maketitle
\begin{abstract}
Surface active agents (surfactants) are common contaminants found on most air-liquid interfaces. Surfactants lower the surface tension, and hence surfactant concentration gradients generate surface tension gradients, which leads to flow. 
A canonical surfactant-induced flow is the outward spreading due to the localised deposition of surfactants onto an otherwise clean liquid interface.
\cite{Eshima26_letter} demonstrated experimentally and theoretically, through a late-time similarity solution, that surfactant solubility enhances surfactant-induced flows of air-liquid-air sheets and that soluble surfactant solutions can be mapped onto equivalent insoluble surfactant solutions.
Here we extend the work theoretically and numerically, developing a systematic framework for the derivation and analysis of such solutions.
The relevant nondimensional parameters are quantitatively defined, and the physically accessible parameter space is wide, which our theory and numerical simulations are able to span.   
We link the solutions for finitely soluble surfactants to the solutions for the limiting regimes of insoluble and infinitely soluble surfactants, where the latter link only holds transiently: the solution of the infinitely soluble limit appears as an intermediate solution that persists ever longer as solubility increases, before the ultimate late-time solution, which maps onto insoluble surfactants, is obtained.
\end{abstract}
\section{Introduction}\label{sec:intro}

Air-liquid-air films appear in many settings across both natural and engineered systems, with fundamental geometries including surface bubble caps, foam films, and the films between coalescing bubbles. In the environment, sea spray aerosols generated from bubble bursting affect the ocean-atmosphere mass exchange \citep{Veron15,Deike22}, while in health, pathogens can be transmitted via aerosols from bubble bursting \citep{Bourouiba21}. In industry, air-liquid-air films, such as foams are key components of personal care, food, and pharmaceutical products. Throughout such applications, the stability of the air-liquid-air sheets is of considerable interest. In the context of surface bubbles, the stability of the bubble cap, an air-liquid-air sheet, determines the bubble lifetime, which affects the bubble bursting process and hence the resulting aerosol distribution. 

Air-liquid interfaces are almost always contaminated, such as by surface active agents (surfactants). Surfactants are molecules that adsorb to the liquid interface and typically lower its surface tension. Surfactants have both anthropogenic and natural origins, where the most familiar everyday example is soap molecules. When there is a concentration gradient of surfactants, there is a surface tension gradient, which leads to a stress known as the \textit{Marangoni} stress and consequently a \textit{Marangoni} flow. Marangoni flows due to surfactants affect a wide range of systems of varying lengthscales and contexts \citep{Fuller12,Manikantan20}, from nanometer-thick soap films \citep{Cantat26} to centimeter- or meter-scale gravity-capillary waves \citep{Liu03,Xu23,Yang26} and plunging breakers \citep{Erinin23}. 

In particular, Marangoni flows in a bubble cap have been identified in the literature as a leading mechanism for surface bubble rupture \citep{Neel18,Poulain18}. More broadly, the stability of air-liquid-air films due to surface tension gradients is of fundamental interest.  
Localised surface tension gradients of air-liquid-air sheets, such as that imposed by the deposition of surfactants, give rise to a Marangoni flow away from the localised region with low surface tension, which leads to a thinning of the sheet. Under the assumption of a thin film, \cite{Bowen13} and \cite{Kitavtsev18} theoretically and numerically analysed the thinning of air-liquid-air films due to Marangoni flow arising from thermal gradients, where it was shown by \cite{Kitavtsev18} that the sheet thinning is exponential in time $t$. \cite{Neel18} showed experimentally that Marangoni flows can thin air-liquid-air films and ultimately lead to rupture. The experiments by \cite{Neel18} imposed localised surface tension gradients through thermal gradients or concentration gradients of solutes that are often thought of as infinitely soluble surfactants (e.g.,  ethanol). Note that mathematically thermal transport is like that of infinitely soluble surfactants. In the opposite limit of insoluble surfactants, still under the assumption of a thin film, late-time similarity solutions show that the film thins proportional to $t^{-1}$ \citep{Eshima25_JFM,Eshima25_PRL}. The transition of $t^{-1}$ to exponential thinning indicates the fundamental role of solubility in setting the strength of surfactant-induced flow in air-liquid-air sheets.

Recently, \cite{Eshima26_letter} demonstrated experimentally that finite solubility enhances surfactant-induced flows of air-liquid-air films by systematically varying the solubility of surfactants, similar to the deposition experiments of \cite{Neel18}.
The physical mechanism for the enhancement is that as the sheet thins, surfactants in the bulk replenish the surface, which sustains a stronger Marangoni flow. 
In particular, the experiments are well described by a physically motivated modification of the insoluble surfactant theory \citep{Eshima25_JFM,Eshima25_PRL}. This enhancement contrasts with the two geometries most studied in the literature of air-liquid-solid films and an air-liquid interface with a deep subphase, where surfactant solubility dampens the Marangoni flow due to the desorption of surfactants into the bulk \citep{Halpern92,Jensen93,Joos85,Banos25}. 

There are three main results in this paper. First, we formulate the problem for surfactants with finite solubility, accounting for the effects of surfactant adsorption, desorption, and cross-film diffusion. We identify the associated non-dimensional parameters that classify how these solubility effects interact. In doing so, we extend the study of \cite{Eshima26_letter} theoretically and numerically across a wide parameter space, building on our earlier framework for insoluble surfactants \citep{Eshima25_JFM,Eshima25_PRL}.    

Second, we show that as for insoluble surfactants, there is a late-time similarity solution, which exhibits a $t^{-1}$ thinning; solubility however alters its prefactor. The prefactor modification is shown to be captured by a single parameter, the non-dimensional depletion length, which denotes a balance between adsorption and desorption.  

Finally, limiting values of solubility are considered to bridge between insoluble, finitely soluble, and infinitely soluble surfactants, where the difference between $t^{-1}$ and exponential film thinning is resolved by intermediate late-time solutions.

The specific outline for the study is described in \S \ref{subsec:outline_study} once we have precisely defined and described the problem and its non-dimensional parameters in \S \S \ref{subsec:tf_eq}, \ref{subsec:surf_volumetric_transport}.

\section{The problem setup}\label{sec:problem_setup}
\subsection{Thin-film equations}\label{subsec:tf_eq}
\begin{figure}
\begin{center}
\includegraphics[width=\textwidth]{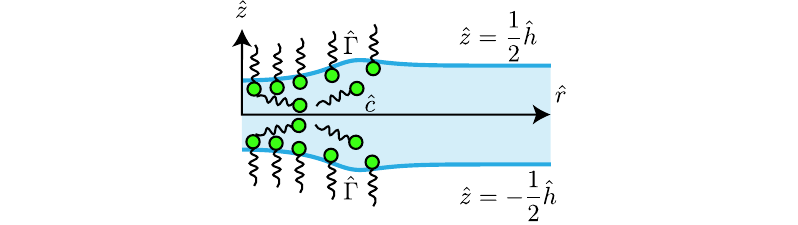}
\end{center}
\caption{Schematic of the top-bottom symmetric axisymmetric sheet of thickness $\hat{h}(\hat{r},\hat{t})$. The $\hat{r}$ axis is in the radial direction and $\hat{z}$ axis is in the axial (cross-film) direction. The top and bottom surfaces of the film are given, respectively, by $\hat{z}=\pm\frac{1}{2}\hat{h}(\hat{r},\hat{t})$. The surface surfactant concentration at the top and bottom is denoted $\hat{\Gamma}$ and the bulk surfactant concentration is denoted $\hat{c}$.}
\label{fig:coord}
\end{figure}

We consider an axisymmetric incompressible Newtonian liquid sheet with viscosity $\mu$ and density $\rho$. We assume top-bottom symmetry of the sheet (for a discussion of top-bottom asymmetry see \cite{Eshima24}). We neglect the effect of the surrounding air. Let $\hat{r}$ and $\hat{z}$ be, respectively, the radial and axial  coordinates, while the radial velocity is denoted $\hat{u}$. Throughout this paper, the term ``cross-film" will be used to describe the $\hat{z}$ direction. The top and bottom of the liquid sheet are given by $\hat{z}=\pm \frac{1}{2}\hat{h}(\hat{r},\hat{t})$ with the surface surfactant concentration given on the top and bottom by $\hat{\Gamma}(\hat{r},\hat{t})$. The variables are summarised in figure \ref{fig:coord}. Also, we assume that the surface tension $\hat{\sigma} = \hat{\sigma}(\hat{\Gamma}(\hat{r},\hat{t}))$ depends solely on the surface surfactant concentration. We denote the constant surface tension without surfactants by 
$\Sigma_c$, which by definition satisfies $\Sigma_c=\hat{\sigma}(0)$. The bulk surfactant concentration is given by $\hat{c}(\hat{r},\hat{z},\hat{t})$. Throughout this paper, the term ``bulk" will be used to refer to the fluid inside the sheet, as opposed to the surface. 

For the initial condition, we consider a sheet otherwise at rest of uniform thickness $\hat{h}_i$ with initial  surface and bulk surfactant concentrations, respectively $\hat{\Gamma}_i(\hat{r})$  and $\hat{c}_i(\hat{r})$, with $\hat{\Gamma}_i(\hat{r}), \hat{c}_i(\hat{r}) \rightarrow 0$ as $\hat{r} \rightarrow \infty$. One can similarly consider initial conditions $\hat{c}_i(\hat{r}, \hat{z})$ that are inhomogeneous in the $\hat{z}$ direction, but such complication will not be necessary for this study; see section \ref{sec:cross-film_diffusion} for details about cross-film inhomogeneity of the bulk surfactant concentration.

Then, the bulk flow equations are given by the axisymmetric Navier-Stokes equations and the boundary conditions are given by standard stress balance conditions along with kinematic conditions, e.g., see the supplementary material of \cite{Eshima25_PRL}. The surface surfactant and bulk surfactant concentrations satisfy advection-diffusion equations. 
We consider linear adsorption and desorption of surfactants, 
such that the flux $\hat{j}$ is given by
\begin{equation}
    \hat{j} = \hat{k}_a \hat{c} - \hat{k}_d \hat{\Gamma},\label{eq:j_flux}
\end{equation}
where $\hat{k}_a$ and $\hat{k}_d$ are, respectively, adsorption and desorption constants. Although it is possible to consider more complicated adsorption and desorption isotherms, in this paper  our mathematical derivations will focus on the late-time limit, $\hat{t}\rightarrow \infty$, and hence the deposited surfactants will spread and at each point $\hat{\Gamma}$ and $\hat{c}$ will tend towards $0$; thus, our final results will not change whether the adsorption is assumed to be linear or not. With the same argument, we will also only consider a linear dependence of surface tension $\hat{\sigma}$ on the surface surfactant concentration $\hat{\Gamma}$.  

The following non-dimensionalisation is consistent with and extends the convention given for insoluble surfactants by \cite{Eshima25_JFM,Eshima25_PRL}. First, consider a characteristic combined amount of surface and bulk surfactant given by $\hat{n}_m := \max \left(\hat{\Gamma}_i + \frac{1}{2}\hat{h}_i\hat{c}_i\right)$. Also, by definition, at kinetic equilibrium (\ref{eq:j_flux}), the flux $\hat{j}$ vanishes. Then, denoting the characteristic surface and bulk surfactant concentrations by $\hat{\Gamma}_c$ and $\hat{c}_c$ respectively, we have that
\begin{equation}
\hat{\Gamma}_c:=\frac{2\Lambda_d\hat{n}_m}{2\Lambda_d + 1}~~~ \hbox{and}~~~ \hat{c}_c:=\frac{2\hat{n}_m}{\left(2\Lambda_d + 1\right)\hat{h}_i}, \label{eq:char_surf}
\end{equation}
where we identify a non-dimensional depletion length (e.g., \cite{Manikantan20})
\begin{equation}
    \Lambda_d := \frac{\hat{k}_a}{\hat{k}_d \hat{h}_i} = \frac{\hat{\Gamma}_c}{\hat{c}_c \hat{h}_i}. \label{eq:def_Lambda_d}
\end{equation}
Physically, $\Lambda_d$ represents the affinity for the surfactant to be on the surface. In other words, $\Lambda_d \gg 1$ means that the surfactant resides mostly on the surface, which is a notion of low solubility, or insolubility, and $\Lambda_d \ll 1$ means that the surfactant resides mostly in the bulk, which is a notion of high solubility. We can now identify a characteristic surface tension deficit, given by 
\begin{equation}
    \Delta \Sigma_c := -\left.\hat{\Gamma}_c \frac{d\hat{\sigma}(\hat{\Gamma})}{d\hat{\Gamma}}\right|_{\hat{\Gamma}=0}. \label{eq:def_surf_deficit}
\end{equation}

With the definition of $\hat{\Gamma}_c$ in (\ref{eq:char_surf}), $\Delta \Sigma_c$ physically represents the maximum surface tension deficit of the system if the surface and bulk surfactant concentrations are in kinetic equilibrium. For more details about the interpretation of $\Delta \Sigma_c$, see \S \ref{subsec:experiment_link}. Letting the conserved (half) total amount of surfactant be given by $N:=2\pi \int_0^{\infty}\hat{r}\left(\hat{\Gamma}_i + \frac{1}{2}\hat{h}_i\hat{c}_i\right)d \hat{r}$, we obtain a characteristic lengthscale, effectively the width of the surfactant distribution, defined as
\begin{equation}
\mathcal{L}_c:=\left(\frac{N}{\pi \hat{n}_m}\right)^{\frac{1}{2}},\label{eq:char_L}
\end{equation}
where the prefactor of $\mathcal{L}_c$ is chosen for analytical convenience in the solution to be identified. 

Finally, we consider the case where both inertia and Marangoni stresses are in the dominant balance and hence $\rho \hat{h}(\partial \hat{u}/\partial \hat{t})\sim (\partial \hat{\sigma}/\partial \hat{r})$. Furthermore, we consider the timescale where appreciable film thinning occurs and hence, from the kinematic boundary conditions and continuity, it follows that $(\partial \hat{h}/\partial \hat{t}) \sim \epsilon \hat{u}$, where $\epsilon :=\hat{h}_i/\mathcal{L}_c$ is the aspect ratio of the dynamics. Combining these scalings, the characteristic Marangoni timescale, as identified by \cite{Neel18}, is given by 
\begin{equation}
    \mathcal{T}_c := \sqrt{\frac{\rho \epsilon \mathcal{L}_c^3}{\Delta \Sigma_c}}.\label{eq:char_T}
\end{equation}

Then, the nondimensional variables (denoted by unhatted variables) satisfy 
\begin{equation}
    (\hat{r}, \hat{t}, \hat{u},\hat{h},\hat{\Gamma}, \hat{c},\hat{\sigma})=\left(\mathcal{L}_c r,\mathcal{T}_ct,\frac{\mathcal{L}_c}{\mathcal{T}_c} u,\epsilon \mathcal{L}_c h,\hat{\Gamma}_c\Gamma,\hat{c}_c c,\Sigma_c+\Delta \Sigma_c \sigma\right).
    \label{eq:nondim}
\end{equation}
Since $\Sigma_c$ is the constant surface tension of the interface without surfactants, $\sigma$ denotes the nondimensional surface tension deficit. The nondimensionalisation means that the following expressions are satisfied:
\begin{subeqnarray} \max\left(\frac{2\Lambda_d\Gamma_i}{2\Lambda_d+1} + \frac{c_i}{2\Lambda_d+1}\right)&=&1,\slabel{eq:max_surf}\\
    \sigma(\Gamma)&=&-\Gamma,\\
    \int_0^{\infty} \left(\frac{2\Lambda_d\Gamma_i}{2\Lambda_d+1} + \frac{c_i}{2\Lambda_d+1}\right)r dr &=& \frac{1}{2}.\slabel{eq:global_conservation_surfactant}
\end{subeqnarray}

The derivation of the thin-film flow equations is standard \citep{Erneux93,DeWit94,Howell96,Brenner99, Breward99,Chomaz01,Savva09,Eshima25_PRL}. By assuming a thin film, $\epsilon \ll 1$, the Navier-Stokes equations including inertia, Marangoni stress, capillary stress, and extensional stress,  along with the continuity equation,  give one-dimensional thin-film equations, where the radial velocity is to leading order one dimensional, $u \approx u(r,t)$:
\begin{subeqnarray}
    \frac{\partial u}{\partial t}+ u\frac{\partial u}{\partial r} &=& -\frac{2}{h}\frac{\partial \Gamma}{\partial r} + \frac{1}{2 \mathcal{M}}\frac{\partial }{\partial r}\left(\frac{1}{r}\frac{\partial}{\partial r}\left(r\frac{\partial h}{\partial r}\right)\right)\nonumber \\
    &+&
   \frac{4}{\textit{Re}} \frac{1}{h}\left(\frac{\partial}{\partial r}\left(\frac{h}{r}\frac{\partial }{\partial r}(r u)\right) - \frac{1}{2} \frac{u}{r}\frac{\partial h}{\partial r}\right),\slabel{eq:tf_mom}\\
    \frac{\partial h}{\partial t}&=&-\frac{1}{r}\frac{\partial }{\partial r}\left(r u h\right).\slabel{eq:tf_mass}\label{eq:tf_flow_total}
\end{subeqnarray}
There are two non-dimensional parameters, where the Marangoni number $\mathcal{M}$ denotes the balance between Marangoni and capillary stresses and the Reynolds number $\textit{Re}$ denotes the balance between inertial and viscous extensional stresses:
\begin{equation}
\mathcal{M}:=\frac{\Delta \Sigma_c}{\epsilon^2\Sigma_c}~,\quad~\textit{Re} := \sqrt{\frac{\rho \Delta \Sigma_c \mathcal{L}_c}{\epsilon \mu^2}}. \label{eq:def_M_Re}
\end{equation}
For example, taking standard properties of water ($\rho \approx 10^3$ kg m $^{-3}$, $\mu \approx 10^{-3}$ Pa s) and an aspect ratio of $\epsilon = 0.1$, along with typical values representative of surfactant deposition $\mathcal{L}_c = 10^{-4}$ m and $\Delta \Sigma_c = 10^{-3}$ N/m (c.f. clean surface tension of water $\approx 7 \times 10^{-2}$ N/m) will lead to a characteristic timescale of $\mathcal{T}_c = 3\times 10^{-4}$ s and $\mathcal{M} \approx 1, \textit{Re}\approx 30$. 

For the examples given in this paper, we will fix the fluid flow non-dimensional parameters as $\mathcal{M}=1$ and $\Rey=10$, since the focus of this study is on the effects of surfactant solubility, rather than the flow parameters. For details of variations due to changes in $\mathcal{M}$ and $\Rey$ for the insoluble limit see \cite{Eshima25_JFM}.

We consider the case where radial surface and bulk P\'eclet numbers, based on diffusivities $D_s$ and $D$, respectively, are large enough that radial diffusion can be neglected. Indeed, for typical surfactant diffusion coefficients $D, D_s = 10^{-10}$ m$^2$ s$^{-1}$ and typical lengthscale $\mathcal{L}_c = 10^{-4}$ m and timescale $\mathcal{T}_c = 10^{-4}$ s, the radial P\'eclet numbers are given by $\mathcal{L}_c^2/\left(D\mathcal{T}_c\right),\mathcal{L}_c^2/\left(D_s\mathcal{T}_c\right) = 10^6 \gg 1$. Then, the surface surfactant equation is a surface advection equation, including exchange with the bulk, given at the interface $z = h/2$ by
\begin{equation}
    \frac{\partial \Gamma}{\partial t}=-\frac{1}{r}\frac{\partial }{\partial r}\left(r u \Gamma\right) + \textit{Bi}(c-\Gamma).\label{eq:tf_surface_surf}
\end{equation}
From continuity, since $u = u(r,t)$, the expression for the velocity in the $z$ direction in the thin film is $-z\left(\partial u/\partial r + u/r\right)$. It follows that the bulk surfactant equation is a volumetric advection equation, including cross-film diffusion, given by
\begin{equation}
    \frac{\partial c}{\partial t}+u\frac{\partial c}{\partial r}-z\left(\frac{\partial u}{\partial r}+\frac{u}{r}\right)\frac{\partial c}{\partial z}= \frac{1}{\textit{Pe}_{\perp}}\frac{\partial^2c}{\partial z^2},\label{eq:tf_bulk_surf_2D}
\end{equation}
where, in the preceding two equations, we identify two non-dimensional parameters that compare surfactant kinetics with the fluid flow,
\begin{equation}
    \textit{Bi}:=\hat{k}_d \mathcal{T}_c~\quad\hbox{and}\quad \quad~\textit{Pe}_{\perp}:=\frac{\hat{h}_i^2}{D \mathcal{T}_c} = \frac{\epsilon^2 \mathcal{L}_c^2}{D\mathcal{T}_c}. \label{eq:def_Bi_Pe}
\end{equation}
Physically, $\textit{Bi}$ compares the desorption timescale to the Marangoni timescale $\mathcal{T}_c$; relatedly, $\textit{Bi}\Lambda_d = \hat{k}_a \hat{h}_i^{-1}\mathcal{T}_c$ compares the adsorption timescale to the Marangoni timescale. The cross-film P\'eclet number $\textit{Pe}_{\perp}$ compares the cross-film diffusion timescale to the Marangoni timescale. 

The initial conditions are given by
\begin{equation}
    u = 0, ~ h = 1, ~\Gamma = \Gamma_i(r), ~c = c_i(r). \label{eq:ic}
\end{equation}
Axisymmetry gives that at $r=0$,
\begin{equation}
    u=\frac{\partial h}{\partial r}=\frac{\partial \Gamma}{\partial r}=\frac{\partial c}{\partial r}=0. \label{eq:axisymmetry}
\end{equation}
Far-field conditions at $r \rightarrow \infty$
are given by
\begin{equation}
    u = 0, ~ h = 1, ~\Gamma = 0, ~c = 0. \label{eq:far_field}
\end{equation}
Due to the assumption of top-bottom symmetry, the boundary condition at $z = 0$ is given by
\begin{equation}
    \left.\frac{\partial c}{\partial z}\right|_{z=0}=0.\label{eq:c_cross-film_bc_z0}
\end{equation}
Balancing the diffusive flux with the kinetic flux (\ref{eq:j_flux}), the boundary condition at $z = \frac{1}{2}h(r,t)$ is given by $\boldsymbol{n} \cdot \nabla c =  -\textit{Pe}_{\perp}\textit{Bi}\Lambda_d (c|_{z=h/2}-\Gamma)$, where $\boldsymbol{n}$ is the unit normal vector directed away from the fluid. Then the leading-order boundary conditions at $z = \frac{1}{2}h(r,t)$ are given by
\begin{equation}
    \left.\frac{\partial c}{\partial z}\right|_{z=h/2}= -\textit{Pe}_{\perp}\textit{Bi}\Lambda_d (c|_{z=h/2}-\Gamma), \label{eq:c_cross-film_bc_interface}
\end{equation}
with corrections at $\textit{O}(\epsilon^2)$ due to the slow variations in the film shape  changing the unit normal $\boldsymbol{n}$.

\subsubsection{The assumption $\textit{Pe}_{\perp}\ll 1$}\label{subsubsec:Pe_perp_assumption}

At this point, we recognize that the P\'eclet number $\textit{Pe}_{\perp}$, defined by (\ref{eq:def_Bi_Pe}), is proportional to $\epsilon^2$, so that $\textit{Pe}_{\perp}\ll 1$ in the limit $\epsilon \rightarrow 0$. In practice, however, the radial P\'eclet number $\mathcal{L}_c^2/(D\mathcal{T}_c)$ is so large that the aspect ratio $\epsilon$ required for $\textit{Pe}_{\perp}\ll 1$ is extremely small. For example, taking typical diffusivity $D = 10^{-10}$ m$^2$ s$^{-1}$ with  $\mathcal{L}_c = 10^{-4}$ m and $\mathcal{T}_c = 10^{-4}$ s requires $\epsilon = 10^{-4}$ in order for  $\textit{Pe}_{\perp}= 10^{-2}$. This corresponds to an initial sheet of thickness $\hat{h}_i = 10^{-8}$ m, at which point van der Waals forces would likely rupture the sheet. The assumption $\textit{Pe}_{\perp} \ll 1$ is therefore not physically realistic in general. 

Nonetheless, we proceed for the time being under the assumption $\textit{Pe}_{\perp}\ll 1$ for mathematical convenience. In \S \ref{sec:cross-film_diffusion}, we demonstrate that this restriction is inconsequential since the late-time similarity solution is unchanged for $\textit{O}(1)$ values of $\textit{Pe}_{\perp}$.

In the limit that $\textit{Pe}_{\perp}\ll 1$, (\ref{eq:tf_bulk_surf_2D}) gives $\partial^2 c/\partial z^2 =0$ to leading order. Then, by symmetry, $c \approx c_0(r,t)$ to leading order and expanding $c(r, z, t) = c_0(r,t) + \textit{Pe}_{\perp} c_1(r,z,t) + \cdots$, we have from (\ref{eq:tf_bulk_surf_2D}) that
\begin{equation}
    \frac{\partial c_0}{\partial t}+ u \frac{\partial c_0}{\partial r}=\frac{\partial^2 c_1}{\partial z^2}.\label{eq:expansion_bulk_transport}
\end{equation}
The boundary condition at $z = 0$ is given by (\ref{eq:c_cross-film_bc_z0}),
\begin{equation}
    \left.\frac{\partial c_1}{\partial z}\right|_{z=0}=0.
\end{equation}
Also, the boundary condition at $z = h/2$ is given by (\ref{eq:c_cross-film_bc_interface}),
\begin{equation}
    \left.\frac{\partial c_1}{\partial z}\right|_{z=\frac{h}{2}}=-\textit{Bi}\Lambda_d (c_0-\Gamma),
\end{equation}
with corrections at $\textit{O}(\epsilon^2 \textit{Pe}_{\perp}^{-1})$, which are small corrections since $ \epsilon^2 \textit{Pe}_{\perp}^{-1}= D\mathcal{T}_c/\mathcal{L}_c^2 \ll 1$.
Then, integrating (\ref{eq:expansion_bulk_transport}) from $z=0$ to $z=h/2$ gives, dropping the subscript for $c \approx c(r,t)$,
\begin{equation}
    \frac{\partial c}{\partial t}+u\frac{\partial c}{\partial r}=-\frac{2}{h}\textit{Bi}\Lambda_d (c-\Gamma).\label{eq:tf_bulk_surf}
\end{equation}

We also give a useful expression that follows from algebraic manipulation of (\ref{eq:global_conservation_surfactant}, \ref{eq:tf_mass}, \ref{eq:tf_surface_surf}, \ref{eq:tf_bulk_surf}), or alternatively from (\ref{eq:global_conservation_surfactant}) directly upon invoking global conservation of surfactant:
\begin{equation}
    \int_0^{\infty} \left(\frac{2\Lambda_d \Gamma}{2\Lambda_d+1} + \frac{hc}{2\Lambda_d+1}\right) r dr = \frac{1}{2}.\label{eq:exact_global_conservation_surfactant}
\end{equation}
In other words, globally no surfactant is gained or lost over time.

\subsubsection{Summary of the thin-film equations}
There are five nondimensional parameters in total:  $\mathcal{M}, \Rey, \textit{Bi}, \Lambda_d, \textit{Pe}_{\perp}$ (\ref{eq:def_Lambda_d}, \ref{eq:def_M_Re}, \ref{eq:def_Bi_Pe}), which are summarised in table \ref{tab:nondimensional}. The nondimensional variables are given by the radial velocity $ u(r,t)$, thickness $ h(r,t)$, surface surfactant concentration $ \Gamma(r,t)$, and bulk surfactant concentration $ c(r,z,t)$. Except for \S \ref{sec:cross-film_diffusion}, we assume that $\textit{Pe}_{\perp}\ll 1$ and hence $c(r,t)$ only (see \S \ref{subsubsec:Pe_perp_assumption}).
Then, the equations for $u,h,\Gamma,c$ are given by (\ref{eq:tf_mom}, \ref{eq:tf_mass}, \ref{eq:tf_surface_surf}, \ref{eq:tf_bulk_surf}) with initial and boundary conditions (\ref{eq:ic}, \ref{eq:axisymmetry}, \ref{eq:far_field}), except for \S \ref{sec:cross-film_diffusion} where we consider (\ref{eq:tf_mom}, \ref{eq:tf_mass}, \ref{eq:tf_surface_surf}, \ref{eq:tf_bulk_surf_2D}) with initial and boundary conditions (\ref{eq:ic}, \ref{eq:axisymmetry}, \ref{eq:far_field}, \ref{eq:c_cross-film_bc_z0}, \ref{eq:c_cross-film_bc_interface}) instead.

The numerical code to solve the thin-film equations will be made available online at \texttt{doi.org/10.34770/145t-as02} \citep{Eshima26_long_PDC}. The supplementary material to this manuscript contains the summary of the numerical code and convergence checks. 

\begin{table}
  \begin{center}
\def~{\hphantom{0}}
  \begin{tabular}{lcccc}
        Name & Symbol   &   Definition & Description & Typical value\\[10pt]
       Marangoni number   & $\mathcal{M}$ & $\frac{\Delta \Sigma_c}{\epsilon^2\Sigma_c}$ & $\frac{\text{Marangoni}}{\text{capillarity}}$ & 1\\[10pt]
       Reynolds number   & $\Rey$ & $\sqrt{\frac{\rho \Delta \Sigma_c \mathcal{L}_c}{\epsilon \mu^2}}$ & $\frac{\text{inertia}}{\text{viscous extensional stress}}$ & 10\\[10pt]
        (nondimensional) Depletion length   & $\Lambda_d$ & $\frac{\hat{k}_a}{\hat{k}_d \hat{h}_i}$ & $\frac{\text{depletion length}}{\text{film thickness}}$ & $ [0,\infty)$\\[10pt]
       Biot number   & \textit{Bi}& $\hat{k}_d \mathcal{T}_c$ & $\frac{\text{desorption}}{\text{advection}}$ & $ [0,\infty)$ \\[10pt]
       (cross-film) Péclet number & $\textit{Pe}_{\perp}$ & $\frac{\hat{h}_i^2}{D \mathcal{T}_c}$ & $\frac{\text{advection}}{\text{cross-film diffusion}}$ &
       $\ll 1$: \S \S \ref{sec:background_solns}, \ref{sec:finitely_soluble_surfactants}, \ref{sec:limiting_behaviours} \\
       
       &&&&$\textit{O}(1)$: \S \ref{sec:cross-film_diffusion}\\[10pt]
  \end{tabular}
  \caption{Summary of the nondimensional parameters considered in this paper.}
  \label{tab:nondimensional}
  \end{center}
\end{table}

\subsubsection{Typical values of the surfactant parameters}\label{subsubsec:surf_params_values}

Here, we discuss the typical values for the surfactant parameters: $\Lambda_d, \textit{Bi}, \textit{Pe}_{\perp}$ (summarised in table \ref{tab:nondimensional}). 

First, we consider the (nondimensional) depletion length $\Lambda_d$. Values of the dimensional depletion length $\hat{k}_a \hat{k}_d^{-1}$ can be readily obtained from equilibrium measurements in the literature by fitting the curve $\hat{k}_a \hat{\Gamma} = \hat{k}_d \hat{c}$. For example, sodium dodecyl sulphate (SDS), which has 12 carbons in its alkyl tail, has $\hat{k}_a \hat{k}_d^{-1} \approx 1 \text{ \textmu m }$ (e.g., from figure 4b of \citep{Varga07}). But $\hat{k}_a \hat{k}_d^{-1}$ varies significantly depending on the surfactant used. For example, sodium octyl sulphate, which has  has 8 carbons in its alkyl tail has $\hat{k}_a\hat{k}_d^{-1} \approx 0.07 \text{ \textmu m }$ (again from figure 4b of \citep{Varga07}) and Triton X-100 has $\hat{k}_a \hat{k}_d^{-1} = \textit{O}(1 \text{ mm})$ \citep{Lin90,Chang95}. Thus, for typical sheet thickness of $\hat{h}_i = \textit{O}(10 \text{ \textmu m})$, $\Lambda_d$ can be large ($\gg\textit{O}(1)$), intermediate ($\textit{O}(1)$), or small ($\ll\textit{O}(1)$).

Next, we consider the Biot number $\textit{Bi}$. Obtaining adsorption and desorption rates $\hat{k}_a$, $\hat{k}_d$ individually require dynamic measurements of surface tension. Dynamic measurements are more difficult than equilibrium measurements which are sufficient to determine only the ratio $\hat{k}_a\hat{k}_d^{-1}$. Then, estimates for $\textit{Bi} = \hat{k}_d\mathcal{T}_c$ for the characteristic timescale $\mathcal{T}_c$ (\ref{eq:char_T}) are less readily available in the literature. For sodium dodecyl sulphate, the desorption rate can be estimated as $\hat{k}_d = \textit{O}(10^3 \text{ s}^{-1})$ \citep{Fernandez05}, which for $\mathcal{T}_c = \textit{O}(100 \text{ \textmu s})$ gives $\textit{Bi}=\textit{O}(0.1)$. As for $\Lambda_d$, different surfactants will lead to greatly different values of $\hat{k}_d$ and hence $\textit{Bi}$ can be large, intermediate, or small.

Finally, we consider the (cross-film) Péclet number $\textit{Pe}_{\perp}$. Taking typical surfactant diffusion as $D = \textit{O}(10^{-10} \text{ m$^2$ s$^{-1}$})$, for film thickness of $\textit{O}(1 \text{ \textmu m})$,  $\textit{Pe}_{\perp}=\hat{h}_i^2/(D\mathcal{T}_c) = \textit{O}(100)$. Thus it is physically relevant to solve for nonzero $\textit{Pe}_{\perp}$, though analytically more simple to solve for $\textit{Pe}_{\perp} \ll 1$.

In summary, the surfactant parameters $\Lambda_d, \textit{Bi}, \textit{Pe}_{\perp}$ varies greatly depending on the surfactant under investigation (and also on the geometry $\hat{h}_i$). Thus, in order to form a complete theory, it is necessary to be able to account for each of these parameters. In this text, having accounted for the wide physical possible ranges of $\Lambda_d, \textit{Bi}, \textit{Pe}_{\perp}$, we show that late time similarity solution ultimately depends only on $\Lambda_d$.

\subsection{Physical intuition: surface and volumetric transport}\label{subsec:surf_volumetric_transport}

In this subsection, we comment on the physical difference between volumetric and surface transport. As described in \S \ref{sec:intro}, it has been shown in previous works that for purely insoluble surfactant transport satisfying
\begin{equation}
    \frac{\partial \Gamma}{\partial t}=-\frac{1}{r}\frac{\partial }{\partial r}\left(r u \Gamma\right), \label{eq:insoluble_surf_transport}
\end{equation}
the minimum thickness of the sheet decays as $t^{-1}$ for $t \gg 1$ \citep{Eshima25_JFM, Eshima25_PRL}. In contrast, for an infinitely soluble surfactant, the bulk remains in equilibrium with the surface, so that $\Gamma=c$ is always satisfied. The transport then satisfies
\begin{subeqnarray}
    \Gamma &=& c,\slabel{eq:inf_soluble_transport_surface}\\
    \frac{\partial c}{\partial t}+u\frac{\partial c}{\partial r}&=&0.\slabel{eq:inf_soluble_transport_bulk}
    \label{eq:inf_soluble_transport}
\end{subeqnarray}
In this limit, the minimum thickness of the sheet decays exponentially with respect to time, which was shown by \cite{Kitavtsev18}, albeit in a slightly different setting, as discussed below in \S \ref{subsec:inf_soluble}.

\begin{figure}
\begin{center}
\includegraphics[width=\textwidth]{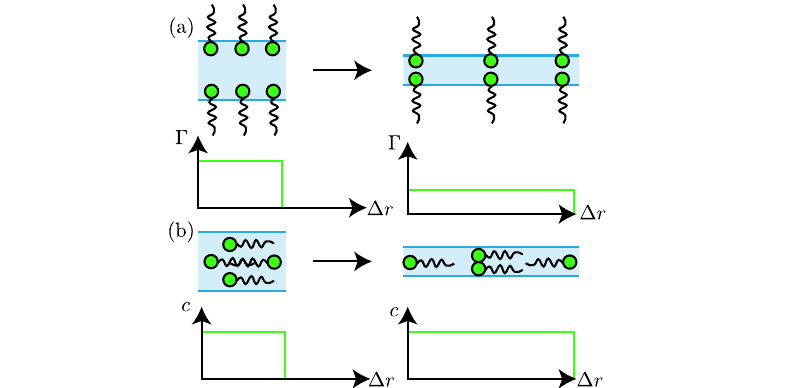}
\end{center}
\caption{Illustration of surface versus volumetric transport for stretching a fluid element of width $\Delta r$, with increasing time illustrated from left to right. (a) Purely surface transport decreases the surface surfactant concentration $\Gamma$. (b) Purely volumetric transport keeps the bulk surfactant concentration $c$ constant.}
\label{fig:surf_vol}
\end{figure}

Upon first inspection, surface advection (\ref{eq:insoluble_surf_transport}) and volumetric advection (\ref{eq:inf_soluble_transport}) look closely related and thus, it may be surprising that there is a significant contrast between the Marangoni thinning due to the two transport equations. However, such a result is physical and intuitive. In figure \ref{fig:surf_vol}, we consider the stretching of a fluid element of width $\Delta r$. Here, the volume is conserved but the surface area increases. Then, for purely interfacial transport as shown in figure \ref{fig:surf_vol}(a), the surface surfactant concentration $\Gamma$ decreases. For purely incompressible volumetric transport as shown in figure \ref{fig:surf_vol}(b), the bulk surfactant concentration $c$ stays constant and hence for infinitely soluble surfactants (\ref{eq:inf_soluble_transport}), the adsorbed surfactant concentration $\Gamma$ remains constant. From this illustration, we see that surface surfactant transport is less efficient at sustaining high surfactant concentrations. Consequently, when insoluble surfactants spread, the Marangoni stress, which is the forcing for the thinning problem, also becomes weaker as time increases, which leads to slower thinning. The same argument holds for 2D Cartesian surface transport $\partial \Gamma/\partial t + \partial (u\Gamma)/\partial x=0$ and 2D Cartesian incompressible volumetric transport $\partial c/\partial t + u\partial c/\partial x=0$.  

For finitely soluble surfactants as considered by (\ref{eq:tf_surface_surf}, \ref{eq:tf_bulk_surf}), the rate of thinning therefore is a balance between volumetric and surface transport. Indeed, as will be shown below, higher solubility surfactants lead to faster film thinning.

\subsection{Outline of the study}\label{subsec:outline_study}

The outline of this paper is as follows. In \S \ref{sec:background_solns}, we summarise the late-time similarity solution of thinning for insoluble surfactants (\S \ref{subsec:insoluble}) and infinitely soluble surfactants (\S \ref{subsec:inf_soluble}), each that have been identified previously in the literature. In \S \ref{sec:finitely_soluble_surfactants}, we derive the late-time similarity solution of thinning for finitely soluble surfactants by extending the derivation for insoluble surfactants. In \S \ref{sec:limiting_behaviours}, we show how the solutions of finitely soluble surfactants are consistent with the solutions of infinitely soluble surfactants via intermediate late-time solutions. Finally \S \ref{sec:cross-film_diffusion} considers cross-film diffusion.  

Throughout \S \S \ref{sec:background_solns} - \ref{sec:cross-film_diffusion}, the equations are nondimensional (see (\ref{eq:nondim})) with the exception of \S \ref{subsec:experiment_link} where connections to experiments are discussed and it is helpful to discuss dimensional quantities.

In particular, we show that the late-time similarity solution of finitely soluble surfactants is analogous to the insoluble case, but with a modified prefactor. The modification is solely through the nondimensional depletion length $\Lambda_d$ and is independent of $\textit{Bi}$ and $\textit{Pe}_{\perp}$.

\section{Background: late-time solution of thinning due to insoluble and infinitely soluble surfactants}\label{sec:background_solns}

\subsection{Insoluble surfactants: a review}\label{subsec:insoluble}

\begin{figure}
\begin{center}
\includegraphics[width=\textwidth]{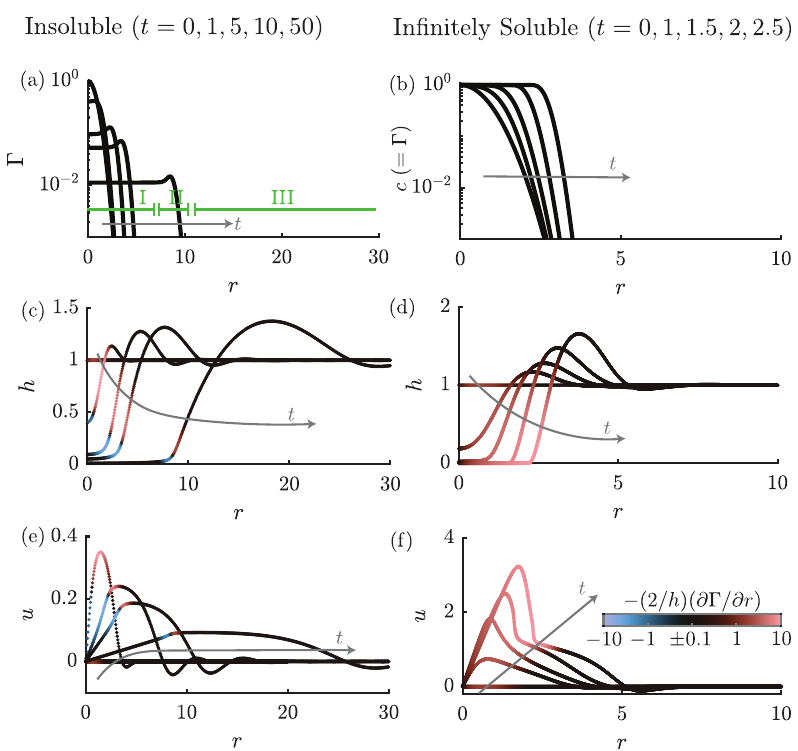}
\end{center}
\caption{Sample evolutions for insoluble and infinitely soluble surfactants. (a,c,e) Sample evolution for insoluble surfactant deposition for $\Gamma_i = e^{-r^2},\mathcal{M}=1,\Rey=10$ (\ref{eq:tf_mom}, \ref{eq:tf_mass}, \ref{eq:insoluble_surf_transport}). (a) Surface surfactant concentration $\Gamma$. (c) Thickness $h$. (e) Horizontal velocity $u$. The times shown are $t = 0,1,5,10,50$. (a,c,e) At late times, there are three regions: region I where there is a spatially uniform $\Gamma$, a transition region II, and a region III where there are no surfactants. (a,c,e) Reproduced from \cite{Eshima25_JFM} (their figure 7), Journal of Fluid Mechanics, Cambridge University Press, under \href{https://creativecommons.org/licenses/}{CC BY}. (b,d,f) Sample evolution for infinitely soluble surfactant deposition for $\Gamma_i = c_i = e^{-r^2},\mathcal{M}=1,\Rey=10$ (\ref{eq:tf_mom}, \ref{eq:tf_mass}, \ref{eq:inf_soluble_transport}). (b) bulk surfactant concentration $c$ (recall by definition, for infinitely soluble surfactants, $\Gamma=c$). (d) $h$. (f) $u$. The times shown are $t = 0,1,1.5,2,2.5$. (c,d,e,f) colours show the Marangoni stress (values within $\pm 0.1$ are set to black). The arrows show the direction of increasing time. The results in all the panels are nondimensional (\ref{eq:nondim}).}
\label{fig:insoluble_infsoluble_sample_dynamics}
\end{figure}

The governing equations for the insoluble surfactant thin-film equations for $u,h$, and $\Gamma$ are given by (\ref{eq:tf_mom}, \ref{eq:tf_mass}, \ref{eq:insoluble_surf_transport}) respectively. Recall that the equations are non-dimensional; see (\ref{eq:nondim}). Numerical solutions are shown in figure \ref{fig:insoluble_infsoluble_sample_dynamics}(a,c,e), which is reproduced from \cite{Eshima25_JFM}. 
As can be seen in figure \ref{fig:insoluble_infsoluble_sample_dynamics}(a), there are three regions at late times: region I where there is a spatially uniform surface surfactant concentration $\Gamma$, region III where there are no surfactants $\Gamma = 0$, and a transition region II. Using asymptotic matching between the regions, \cite{Eshima25_JFM,Eshima25_PRL} identified the late-time similarity solution. 

\begin{figure}
\begin{center}
\includegraphics[width=\textwidth]{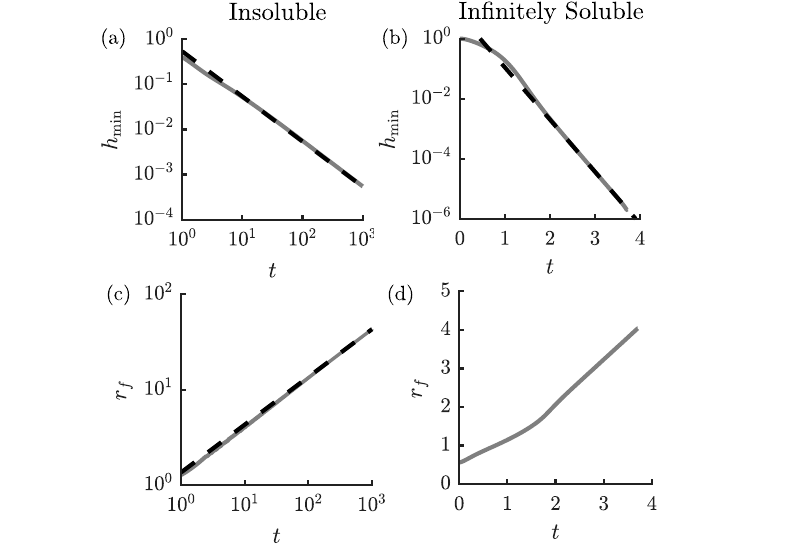}
\end{center}
\caption{Temporal evolution of the minimum thickness $h_{\text{min}}$ and surfactant front location $r_f$ for insoluble and infinitely soluble surfactants. (a,c) Comparison of the predicted similarity solution (solid curves) with the thin-film equations (\ref{eq:tf_mom}, \ref{eq:tf_mass}, \ref{eq:insoluble_surf_transport}) (dashed lines) for insoluble surfactant deposition with $\Gamma_i = e^{-r^2},\mathcal{M}=1,\Rey=10$. There are no fitting parameters in (a,c). (a) Time evolution of $h_{\text{min}}$. (c) Time evolution of front location $r_f$. (b,d) Numerical solution to the thin-film equations (\ref{eq:tf_mom}, \ref{eq:tf_mass}, \ref{eq:inf_soluble_transport}) for infinitely soluble surfactant deposition with $\Gamma_i=c_i = e^{-r^2},\mathcal{M}=1,\Rey=10$. (b) Time evolution of $h_{\text{min}}$, where the analytical prediction $h_{\text{min}}\sim e^{-4t}$ is shown as a dashed line (more precisely, the line is $h_{\text{min}}= 6e^{-4t}$ where the prefactor $6$ is a fit). (d) Time evolution of $r_f$, which suggests $r_f \sim t$ at late times $t\gg 1$. The results in all the panels are nondimensional (\ref{eq:nondim}).}
\label{fig:insoluble_inf_soluble_verification}
\end{figure}

In summary, the minimum thickness $h_{\text{min}}(t)$ and the moving surfactant front location $r_f(t)$ satisfy
\begin{equation}
h_{\text{min}}(t)=\eta_f^{-2}t^{-1} \text{ and }r_f(t)=\eta_ft^{1/2},
\end{equation}
where $\eta_f$ is a function of $\mathcal{M}$ and $\Rey$ that can be evaluated numerically from the ordinary differential equations associated with the similarity solution, i.e., it is not a fitting parameter. A comparison between the similarity solution and the numerical solution of the thin-film equations is shown in figure \ref{fig:insoluble_inf_soluble_verification}(a,c). 

\subsection{Infinitely soluble surfactants}\label{subsec:inf_soluble}

The governing equations for the infinitely soluble surfactant thin-film equations for $u,h$, and $c$ are given by (\ref{eq:tf_mom}, \ref{eq:tf_mass}, \ref{eq:inf_soluble_transport}) respectively. Recall that the equations are non-dimensional; see (\ref{eq:nondim}). Numerical solutions are shown in figure \ref{fig:insoluble_infsoluble_sample_dynamics}(b,d,f), which shows several distinguishing features of the dynamics due to solubility in comparison to insoluble surfactants (figure \ref{fig:insoluble_infsoluble_sample_dynamics}(a,c,e)). First, the surface surfactant concentration in the region with the thinnest film remains $\textit{O}(1)$, as highlighted in \S \ref{subsec:surf_volumetric_transport}. Additionally, the width of the surfactant front is much more narrow than the insoluble case, which is best seen by the region with sharp spatial gradient of $u$ in  figure \ref{fig:insoluble_infsoluble_sample_dynamics}(f). These two observations together lead to much stronger Marangoni stress than for the case of insoluble surfactants. Consequently, there is a larger radial velocity $u$ (figure \ref{fig:insoluble_infsoluble_sample_dynamics}(f)) and hence much faster thinning of the sheet (figure \ref{fig:insoluble_infsoluble_sample_dynamics}(d)).

\cite{Bowen13} and \cite{Kitavtsev18} considered similar sets of equations to (\ref{eq:tf_mom}, \ref{eq:tf_mass}, \ref{eq:inf_soluble_transport}), as they were considering thin-film evolution due to thermal gradients and corresponding Marangoni-driven flows. Temperature fields are mathematically  equivalent to infinitely soluble surfactants. Furthermore, \cite{Kitavtsev18} showed that the resulting Marangoni thinning is exponential with respect to time and hence does not formally produce a finite-time pinch-off. There are, however, two notable differences to the set of equations considered by \cite{Kitavtsev18} and that considered by  us (\ref{eq:tf_mom}, \ref{eq:tf_mass}, \ref{eq:inf_soluble_transport}). Firstly, their domain is Cartesian, 2D, and periodic in the in-plane coordinate, denoted by $x$, whereas we consider a 2D axisymmetric domain with far-field boundary conditions with in-plane coordinate $r$. Secondly, their thermal (or infinitely soluble surfactant) transport equation contains an in-plane diffusion term, which we neglect (see discussions surrounding (\ref{eq:tf_surface_surf}, \ref{eq:tf_bulk_surf_2D})). This difference is understandable since the thermal diffusion coefficient, e.g., in water of order $\textit{O}(10^{-7} \text{ m$^2$ s$^{-1}$})$, is much larger than the surfactant diffusion coefficient, which is typically order $\textit{O}(10^{-10} \text{ m$^2$ s$^{-1}$})$. 
Furthermore, accounting for the in-plane diffusion term is important for thermal (or infinitely soluble surfactant) deposition, where strong Marangoni forcing narrows the surfactant front region width $\Delta r$. 
Indeed, for the 2D periodic case, \cite{Kitavtsev18} showed that the surfactant front has an exponentially small width. In comparison, for the insoluble surfactant case, and consequently the finitely soluble surfactant case, as will be shown in \S \ref{sec:finitely_soluble_surfactants}, the Marangoni forcing across the surfactant front (region II) is comparatively much weaker and hence in-plane diffusion is less important. For example, consider the initial surfactant distribution $\Gamma_i(r) \sim r^{-\alpha}$ as $r \rightarrow \infty$ with $\alpha >2$, i.e., values of $\alpha$ corresponding to a finite amount of surfactants. Then, region II has width $\textit{O}\left(t^{(2\alpha -2)^{-1}}\right)$ \citep{Eshima25_JFM}, which is $\textit{O}(1)$ or greater; when the initial surfactant distribution decays faster than polynomial, the expression corresponds to $\alpha = \infty$. Since we consider finitely soluble surfactants, we therefore neglect in-plane diffusion throughout this paper.

Since the infinitely soluble limit is not the focus of this paper, we do not derive the corresponding late-time similarity solution. We expect that taking the same approach as \cite{Kitavtsev18} would allow for such a similarity solution to be found. However, for a Gaussian deposition $\Gamma_i = c_i = e^{-r^2}$, we find that it is possible to deduce that the minimum thickness $h_{\text{min}}\sim e^{-4t}$ at late times without obtaining the full similarity solution (see Appendix \ref{app:inf_soluble_derivation}).

The temporal evolutions of $h_{\text{min}}$ and $r_f$ are shown in figure \ref{fig:insoluble_inf_soluble_verification}(b,d). The dashed line in figure \ref{fig:insoluble_inf_soluble_verification}(b) shows the analytical prediction $h_{\text{min}} \sim e^{-4t}$. Moreover, the surfactant front $r_f$ appears to scale like $t$ at late times, instead of $t^{1/2}$ as in the insoluble case. An increase in the exponent for the infinitely soluble limit makes sense physically due to the stronger $\textit{O}(1)$ jump in Marangoni stress across the surfactant front. 

The infinitely soluble scaling $r_f \sim t$ can be seen as follows. By regarding the surfactant spreading as a propagating shock, Rankine-Hugoniot jump conditions follow from the momentum equation by balancing the flux due to Marangoni stress $\partial \Gamma/\partial r$ and advection terms $\partial (h u^2)/\partial r$ (\ref{eq:tf_mom}), which gives that the surfactant concentration jump across the front satisfies $\Delta \Gamma \sim (dr_f/dt)^2$ and hence $r_f \sim t$.
However, we were not able to run the numerical simulations longer in our code to numerically confirm this scaling. A code specifically designed to handle exponential behaviour, such as that developed by \cite{Kitavtsev18}, should be able to give numerical solutions at later times.    

\section{Late-time similarity solution of thinning due to finitely soluble surfactants}\label{sec:finitely_soluble_surfactants}

In this section, we derive the late-time similarity solution associated with finitely soluble surfactants. The thin-film equations we consider in this section for $(u,h,\Gamma,c)$ are given by (\ref{eq:tf_mom}, \ref{eq:tf_mass}, \ref{eq:tf_surface_surf}, \ref{eq:tf_bulk_surf}). The equations in this section are non-dimensional (see (\ref{eq:nondim})), except for \S \ref{subsec:experiment_link} where connections to experiments are discussed and dimensional equations are more natural.

A representative numerical solution is shown in figure \ref{fig:soluble_sample_dynamics}. In panel (a), at late times, $t \gg 1$, the magnitude of the jump in surface surfactant concentration $\Gamma$ at the front decreases over time. Thus, finitely soluble surfactant deposition at late times is similar to insoluble surfactant deposition rather than infinitely soluble surfactant deposition (figure \ref{fig:insoluble_infsoluble_sample_dynamics}(a,b)). In other words, surface transport dominates volumetric transport at late times (figure \ref{fig:surf_vol}). Another observation that can be made from the results in figure \ref{fig:soluble_sample_dynamics}(a,b) is that $\Gamma \approx c$ at late times $t \gg 1$. 

\subsection{The derivation of the late-time similarity solution}\label{subsec:deriv_simil_soln_fin_soluble}

\begin{figure}
\begin{center}
\includegraphics[width=\textwidth]{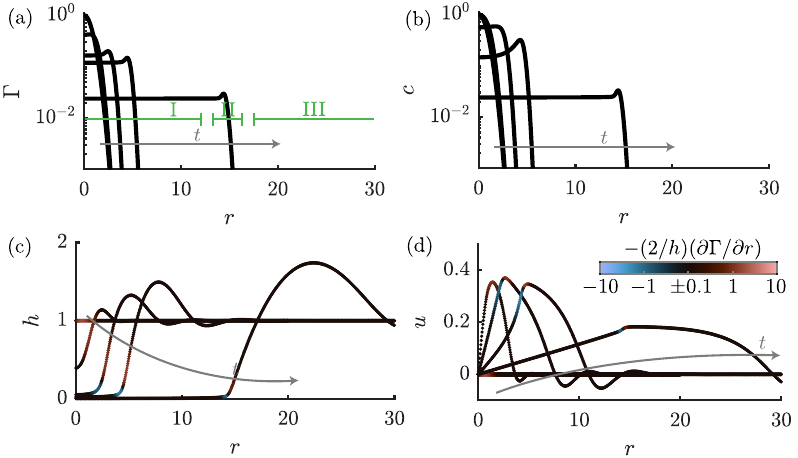}
\end{center}
\caption{Sample evolution for finitely soluble surfactant deposition for $\Gamma_i=c_i = e^{-r^2}, \mathcal{M}=1, \Rey=10, \textit{Bi}=0.1$, and $\Lambda_d = 0.1$ (\ref{eq:tf_mom}, \ref{eq:tf_mass}, \ref{eq:tf_surface_surf}, \ref{eq:tf_bulk_surf}). (a) Surface surfactant concentration $\Gamma$. (b) Bulk surfactant concentration $c$. (c) Thickness $h$. (d) Horizontal velocity $u$. For (c,d), colours show the Marangoni stress (values within $\pm 0.1$ are set to black). The times shown are $t = 0,1,5,10,50$ and the arrows show the direction of increasing time. At late times, as for insoluble surfactants, there are three regions: region I where there is a spatially uniform $\Gamma$, a transition region II, and a region III where there are no surfactants. The results in all the panels are nondimensional (\ref{eq:nondim}).}
\label{fig:soluble_sample_dynamics}
\end{figure}

As for insoluble surfactant deposition, there are three regions: region I, where there is a spatially uniform concentration of surfactants, a transition region II, and region III where there are no surfactants (see figure \ref{fig:soluble_sample_dynamics}(a)). The prefactor of the front location $\eta_f$ is found once the solutions for the different regions have been matched. In this subsection, subscripts are used to emphasise the region under discussion. 

The key step for the inclusion of finite solubility is as follows. We focus on region I. 
At late times $t \gg 1$, balancing the time derivative and advection terms of (\ref{eq:tf_surface_surf}), which have the scale $\Gamma_{\text{I}}/t$, with the adsorption-desorption term $\textit{Bi}(c_{\text{I}}-\Gamma_{\text{I}})$ yields $\Gamma_{\text{I}}= c_{\text{I}}(1+\textit{O}(t^{-1}))$ and hence $\Gamma_{\text{I}}= c_{\text{I}}$ to leading order. In other words, kinetic equilibrium is reached to leading order for $t \gg 1$. Substituting $\Gamma_{\text{I}}= c_{\text{I}}$ and $h_{\text{I}}\ll 1$ into global conservation of surfactant (\ref{eq:exact_global_conservation_surfactant}) gives 
\begin{equation}
    \int_0^{\infty} \frac{2\Lambda_d \Gamma_{\text{I}}}{2\Lambda_d+1} r dr=\frac{1}{2}. \label{eq:soluble_late_times_conservation_surfactant}
\end{equation}
A physical interpretation of (\ref{eq:soluble_late_times_conservation_surfactant}) is that at late times, most of the surfactants are on the surface. 

The rest of the derivation of the similarity solution is the same as for the insoluble case (see \cite{Eshima25_JFM, Eshima25_PRL}), with the only difference appearing in the total amount of surfactant, which satisfies (\ref{eq:soluble_late_times_conservation_surfactant}) rather than $\int_0^{\infty}\Gamma_{\text{I}}r dr = 1/2$, which is the case for the insoluble limit $\Lambda_d = \infty$. Thus, the details, such as the scaling arguments for the dominant balance in each region are omitted. 

Region I is the range $ 0 \leq r< r_f:= \eta_f t^{1/2}$, where the prefactor $\eta_f$ is to be determined
once the solutions of the different regions have been matched. The similarity coordinate is then defined by $\eta := \eta_f^{-1} rt^{-1/2}$. Scalings in the conservation of momentum (\ref{eq:tf_mom}) gives $\partial \Gamma_{\text{I}}/\partial r = 0$. Then, using $\Gamma_{\text{I}}=c_{\text{I}}$, along with (\ref{eq:tf_mass}, \ref{eq:tf_surface_surf})  and setting prefactors via (\ref{eq:soluble_late_times_conservation_surfactant}) leads to   
\begin{equation}
    u_{\text{I}}=\frac{1}{2}\eta_f t^{-\frac{1}{2}}\eta, \quad h_{\text{I}}=\eta_f^{-2}t^{-1}f(\eta), \quad \Gamma_{\text{I}}=c_{\text{I}}=\frac{2\Lambda_d+1}{2\Lambda_d}\eta_f^{-2}t^{-1},\label{eq:sol_regI}
\end{equation}
where $\eta \in [0,1)$ and, as detailed in Appendix \ref{app:horiz_Lag_coords}, transformation to radial Lagrangian coordinates shows that the function $f(\eta)$ satisfies
\begin{equation}
    1 = \left(\frac{2\Lambda_d}{2\Lambda_d +1}\Gamma_i(s(\eta)) + \frac{1}{2\Lambda_d+1}c_i(s(\eta))\right)f(\eta),\label{eq:f_regI_finite_soluble}
\end{equation}
where $s(\eta) = \left(2\int_0^{\eta} \eta'f(\eta') d \eta'\right)^{\frac{1}{2}}$. From the nondimensionalisation (\ref{eq:max_surf}), it then follows that $h_{\text{min}}=\eta_f^{-2}t^{-1}$.

Region II is the transition region, located at $r \approx \eta_f t^{1/2}$ with spatial coordinate $\Delta r_{\text{II}} := r-\eta_f t^{1/2}$. Scalings in the conservation of momentum (\ref{eq:tf_mom}) give that the dominant balance is between Marangoni and capillary stresses in region II. Then, integrating the Marangoni and capillary stress balance, while matching the value of $\Gamma_{\text{II}}$ onto the value of $\Gamma_{\text{I}}$ in (\ref{eq:sol_regI}) gives
\begin{equation}
    \Gamma_{\text{II}}-\frac{2\Lambda_d+1}{2\Lambda_d}\eta_f^{-2}t^{-1}=\frac{1}{4\mathcal{M}}\left(h_{\text{II}}\frac{\partial^2 h_{\text{II}}}{\partial (\Delta r_{\text{II}})^2}-\frac{1}{2}\left(\frac{\partial h_{\text{II}}}{\partial \Delta r_{\text{II}}}\right)^2\right).\label{eq:soluble_regII}
\end{equation}
Then, the matching condition to region III is deduced to be
\begin{equation}
    -\frac{2\Lambda_d+1}{2\Lambda_d}\eta_f^{-2}t^{-1}=\lim_{\text{II}\rightarrow \text{III}}\frac{1}{4\mathcal{M}}\left(h\frac{\partial^2 h}{\partial (\Delta r_{\text{II}})^2}-\frac{1}{2}\left(\frac{\partial h}{\partial \Delta r_{\text{II}}}\right)^2\right),\label{eq:soluble_regII_to_III}
\end{equation}

Region III is where there are no surfactants, and is located at the range $r>\eta_f t^{1/2}$. Scaling arguments lead to the self-similar ansatz 
\begin{equation}
u_{\text{III}}=\eta_ft^{-\frac{1}{2}}U(\eta), ~h_{\text{III}}=H(\eta),\label{eq:region_III_ansatz}
\end{equation}
where $\eta \in (1,\infty)$. Then,  (\ref{eq:tf_mom}, \ref{eq:tf_mass}) turn into a system of first-order ordinary differential equations (ODEs). The ODEs are exactly the same as for the insoluble surfactant case, with the sole modification arising from the factor $(2\Lambda_d+1)/(2\Lambda_d)$ in (\ref{eq:soluble_regII_to_III}). Defining $(J,K):=(dH/d\eta, d^2H/d\eta^2)$, the ODEs are given by
\begin{subeqnarray}
    \frac{dU}{d\eta} &=& \left(\frac{\eta}{2}-U\right)\frac{J}{H}-\frac{U}{\eta}, \slabel{eq:U_similsoln}\\
    \frac{dH}{d\eta}&=&J,\slabel{eq:H_similsoln}\\
    \frac{dJ}{d\eta} &=& K, \slabel{eq:J_similsoln}\\
    \frac{dK}{d\eta} &=& -\frac{K}{\eta}+\frac{J}{\eta^2}+2\eta_f^4\mathcal{M}\left(-\frac{U}{2}+\left(U-\frac{\eta}{2}\right)\left(\left(\frac{\eta}{2}-U\right)\frac{J}{H}-\frac{U}{\eta}\right)\right) \nonumber\\
    &-&\frac{8\eta_f^2\mathcal{M}}{\Rey }\left(\left(\frac{1}{2}+\frac{U}{2\eta}-\left(\frac{\eta}{2}-U\right)\frac{J}{H}\right)\frac{J}{H} +\left(\frac{\eta}{2}-U\right)\frac{K}{H}\right). \slabel{eq:K_similsoln}
    \label{eq:soluble_regIII}
\end{subeqnarray}
The boundary conditions at $\eta = 1+ \delta$ for $\delta \ll 1$ are derived by matching region III to region II:
\begin{subeqnarray}
    U(1+\delta) &=& \frac{1}{2} +\dots,\slabel{eq:soluble_U_BC}\\
    H(1+\delta) &=& \sqrt{8\mathcal{M}\left(\frac{2\Lambda_d+1}{2\Lambda_d}\right)}\delta - \frac{\eta_f^2\mathcal{M}}{\Rey }\delta^2 \log \delta + q\delta^2 +\dots,\slabel{eq:soluble_H_BC}
    \label{eq:soluble_regIII_BC}
\end{subeqnarray}
for some constant  $q$. It is worth noting from (\ref{eq:soluble_regIII}, \ref{eq:soluble_regIII_BC}) that the late-time similarity solution is therefore independent of $\textit{Bi}$  and the modification due to solubility occurs solely through the parameter $\Lambda_d$. 

There are two constraints for the system of ODEs. First, the sheet is undisturbed in the far field. Second, by global conservation of mass, the mass that was originally in region I at $t = 0$ is in region III at late times. Thus, the similarity solution can be found by a shooting algorithm, where $\eta_f$ and $q$ are adjusted so that two constraints are satisfied  
\begin{equation}
    H(\infty)= 1 \text{ and } \int_1^{\infty}(H-1)\eta d\eta=\frac{1}{2}.\label{eq:shooting_constraints}
\end{equation}

A mathematical observation of the system of ODEs (\ref{eq:soluble_regIII}, \ref{eq:soluble_regIII_BC}, \ref{eq:shooting_constraints}) is that the four constants $\eta_f, \mathcal{M},\Rey, \Lambda_d$ only appear in three expressions $\eta_f^4 \mathcal{M}$, $\eta_f^2\mathcal{M}\Rey^{-1}$, and $\mathcal{M}(2\Lambda_d+1)/(2\Lambda_d)$. Then, appropriate change of variables of $\eta_f, \mathcal{M},\Rey, \Lambda_d$ leaves the solution to the system of ODEs (\ref{eq:soluble_regIII}, \ref{eq:soluble_regIII_BC}, \ref{eq:shooting_constraints}) unchanged. One such change of variables is 
\begin{equation}
    (\tilde{\eta}_f, \tilde{\mathcal{M}}, \tilde{\Rey}, \tilde{\Lambda_d}) := \left(\eta_f \left(\frac{2\Lambda_d+1}{2\Lambda_d}\right)^{-1/4},\mathcal{M}\left(\frac{2\Lambda_d+1}{2\Lambda_d}\right), \Rey \left(\frac{2\Lambda_d+1}{2\Lambda_d}\right)^{1/2}, \infty\right),
\end{equation}
which satisfies $\tilde{\eta}_f^4 \tilde{\mathcal{M}}= \eta_f^4 \mathcal{M}$, $\tilde{\eta}_f^2\tilde{\mathcal{M}}\tilde{\Rey}^{-1} = \eta_f^2\mathcal{M}\Rey^{-1}$, $\tilde{\mathcal{M}}(2\tilde{\Lambda_d}+1)/(2\tilde{\Lambda_d}) = \mathcal{M}(2\Lambda_d+1)/(2\Lambda_d)$. Thus, the following identity is satisfied exactly for the front prefactor function $\eta_f(\mathcal{M},\Rey,\Lambda_d)$:
\begin{equation}
    \eta_f(\mathcal{M},\Rey,\Lambda_d) = \left(\frac{2\Lambda_d+1}{2\Lambda_d}\right)^{1/4} \eta_f\left(\mathcal{M}\left(\frac{2\Lambda_d+1}{2\Lambda_d}\right), \Rey \left(\frac{2\Lambda_d+1}{2\Lambda_d}\right)^{1/2},\infty\right).\label{eq:eta_f_soluble}
\end{equation}
Equation (\ref{eq:eta_f_soluble}) shows mathematically that deposition and transport of finitely soluble surfactant  is like that of deposition and transport of insoluble surfactant, $\Lambda_d = \infty$,  but modified by scalings involving $\Lambda_d$. Furthermore, it follows directly from (\ref{eq:eta_f_soluble})  that the finitely soluble similarity solution is consistent with the similarity solution for insoluble surfactants, since $\lim_{\Lambda_d \rightarrow \infty} \eta_f(\mathcal{M},\Rey,\Lambda_d) = \eta_f(\mathcal{M},\Rey,\infty)$. 

In Appendix \ref{app:lim_highM_Re}, the system of ODEs (\ref{eq:soluble_regIII}, \ref{eq:soluble_regIII_BC}, \ref{eq:shooting_constraints}) is solved analytically for limiting values of $(\mathcal{M},\Rey, \Lambda_d)$, which leads to the following limit 
\begin{equation}
    \lim_{\alpha \rightarrow \infty}\eta_f\left(\mathcal{M}\alpha, \Rey \alpha^{1/2}, \infty\right)=2.
\end{equation}
Then, it follows from (\ref{eq:eta_f_soluble}) that for the highly soluble limit $\Lambda_d \ll 1$, the late-time similarity solution becomes independent of $\mathcal{M}$ and $\Rey$ and satisfies
\begin{equation}
    \eta_f(\mathcal{M},\Rey,\Lambda_d) \approx 2^{3/4}\Lambda_d^{-1/4}.\label{eq:eta_f_Ld_small_limit}
\end{equation}
Thus, $\eta_f$ becomes singular in the limit $\Lambda_d \rightarrow 0^+$, which suggests a change in the scaling law at $\Lambda_d = 0$ as discussed previously in \S \ref{subsec:inf_soluble}, where there was exponential decay. Thus, it is difficult to see from (\ref{eq:eta_f_Ld_small_limit}) alone that there is consistency with infinitely soluble surfactants for $0<\Lambda_d \ll 1$. In \S \ref{sec:limiting_behaviours}, we show consistency with infinitely soluble surfactants through intermediate late-time behaviour.

\subsection{Physical interpretation of the solution}

\begin{figure}
\begin{center}
\includegraphics[width=\textwidth]{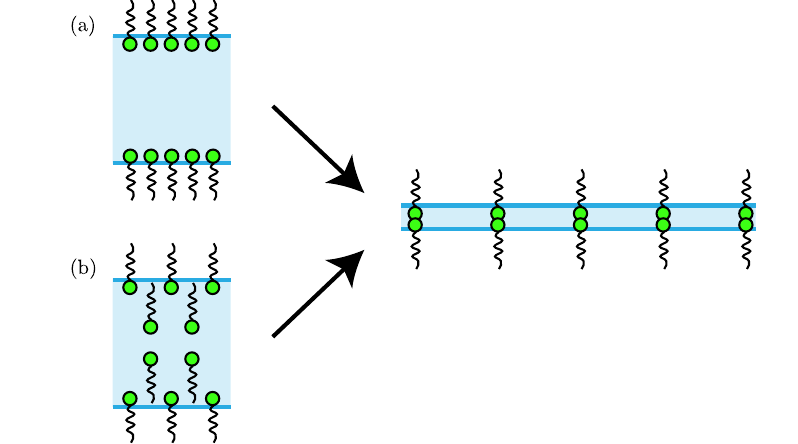}
\end{center}
\caption{Intuition for the late-time similarity solution of finitely soluble surfactants. (a) Fluid element with insoluble surfactants. (b) Fluid element with finitely soluble surfactants with the same amount of fluid and surfactants as (a). For (b), upon stretching the fluid element extremely thin and letting the surfactants come to kinetic equilibrium, most of the surfactants by count will be on the surface and hence the fluid element is like that of (a) when similarly stretched.}
\label{fig:intuition_similarity_solution}
\end{figure}

Here, we describe a physical interpretation for the relationship of the solution for finitely soluble and insoluble surfactants (\ref{eq:eta_f_soluble}), which is shown schematically in figure \ref{fig:intuition_similarity_solution}.
Consider (a) a fluid element with insoluble surfactants and (b) another fluid element with finitely soluble surfactants, where the combined number of surface and bulk surfactants and the total amount of fluid is the same. For (b), if the fluid element is stretched extremely thin and the surfactants are allowed to come to kinetic equilibrium so that $\Gamma = c$, most of the surfactants by count will be on the surface since $c$ is volumetric. The resulting state is like that of (a) when similarly stretched. 

For the problem of surfactant deposition, the physical intuition is the same. At late times, since the sheet is thin in the region with surfactants, most of the surfactants by count are on the surface and hence the profile of surface surfactant concentration $\Gamma$ is as if the surfactants were insoluble and started on the surface, which in turn sets the same Marangoni stress and hence the same surfactant front propagation. Indeed, for a given initial condition $\hat{\Gamma}_i$ and $\hat{c}_i$ for finitely soluble surfactants, consider the effective insoluble surfactant configuration ($\Lambda_d = \infty$) with all the surfactants on the surface $\hat{\Gamma}_{i,\text{eff}} = \hat{\Gamma}_i+\frac{1}{2}\hat{h}_i \hat{c}_i$, $\hat{c}_{i,\text{eff}}=0$. Then, the dimensional characteristic scales identified in (\ref{eq:def_surf_deficit}, \ref{eq:char_L}, \ref{eq:char_T}) and flow parameters (\ref{eq:def_M_Re}) are
\begin{equation}
    (\Delta \Sigma_{c,\text{eff}}, \mathcal{L}_{c,\text{eff}}, \mathcal{T}_{c,\text{eff}}, \mathcal{M}_{\text{eff}},\Rey_{\text{eff}})=\left(\frac{2\Lambda_d+1}{2\Lambda_d}\Delta \Sigma_c, \mathcal{L}_c, \sqrt{\frac{2\Lambda_d}{2\Lambda_d+1}}\mathcal{T}_c,\frac{2\Lambda_d+1}{2\Lambda_d} \mathcal{M}, \sqrt{\frac{2\Lambda_d+1}{2\Lambda_d}}\Rey \right),\label{eq:effective_st_transform}
\end{equation} 
which recovers the rescaling identified in (\ref{eq:eta_f_soluble}). 

\subsection{Verification of the solution and results}\label{subsec:fin_sol_verif}

In this section, we verify the late-time similarity solution identified in this section for finitely soluble surfactants. The similarity solution prediction for the minimum thickness $h_{\text{min}}=\eta_f^{-2}t^{-1}$ (dashed lines) is compared against the numerical solutions of the thin-film equations (solid curves) in figure \ref{fig:finite_soluble_verification}(a) for the case $\Gamma_i = c_i = e^{-r^2}$; in Appendix \ref{app:different_ic}, an example with $\Gamma_i \neq c_i$ is shown. Cases for $\Lambda_d = 0.1,1,10$ for the numerical solutions are shown, respectively, in pink, blue, and purple ($\textit{Bi}=0.1,1,10$ are dark-shade, medium-shade, light-shade). The analogous plot but for the surfactant front location $r_f = \eta_f t^{1/2}$ is shown in figure \ref{fig:finite_soluble_verification}(d). The thickness profiles $h$ versus $rt^{-1/2}$ at a late time ($t=1000$ shown here) are shown in figure \ref{fig:finite_soluble_verification}(b,c,e). We see that the late-time similarity solution agrees well with the numerical solution at late times without any fitting parameters. 

Furthermore, we observe two physical predictions about the effect of solubility on the late-time similarity solution. First, the effect of solubility at late times can be solely described by $\Lambda_d$ and is independent of $\textit{Bi}$, as can be seen from different $\textit{Bi}$ collapsing onto the same profiles at late time (figure \ref{fig:finite_soluble_verification}(b,c,e)). Secondly, smaller $\Lambda_d$, i.e., more soluble,  leads to faster thinning and front propagation, as can be seen by the decrease of $h_{\text{min}}$ in figure \ref{fig:finite_soluble_verification}(a) and the increase of $r_f$ in figure \ref{fig:finite_soluble_verification}(d) as $\Lambda_d$ decreases.

\begin{figure}
\begin{center}
\includegraphics[width=\textwidth]{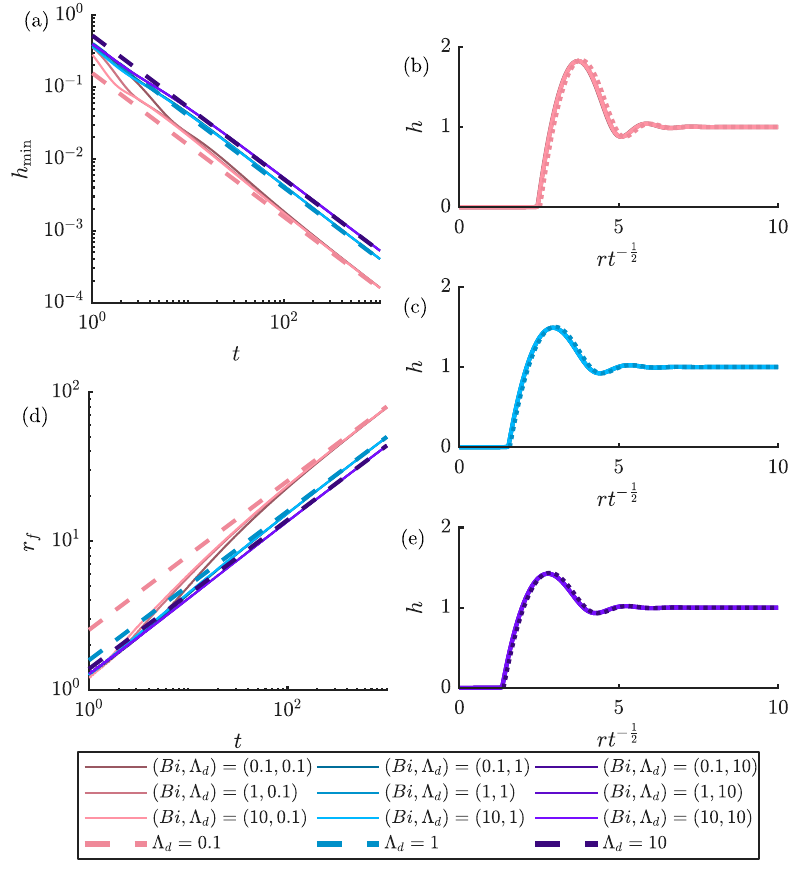}
\end{center}
\caption{Deposition of finitely soluble surfactants with $\Gamma_i=c_i = e^{-r^2}, \mathcal{M}=1$, and $\Rey=10$. Dashed lines show the similarity solution prediction (\ref{eq:soluble_regIII}, \ref{eq:soluble_regIII_BC}) and the  solid curves show the numerical solutions of the thin-film equations (\ref{eq:tf_mom}, \ref{eq:tf_mass}, \ref{eq:tf_surface_surf}, \ref{eq:tf_bulk_surf}). The numerical solutions are coloured according to $\Lambda_d = 0.1, 1, 10$ (pink, blue, purple) and shaded according to $\textit{Bi}=0.1,1,10$ (dark, medium, light). The similarity solutions are also coloured according to $\Lambda_d = 0.1,1,10$; recall that the similarity solutions are independent of $\textit{Bi}$. (a) Evolution of minimum thickness $h_{\text{min}}$ over time $t$. (d) Evolution of the surfactant front location $r_f$ over time $t$. (b,c,e) The thickness profile $h$ against $rt^{-1/2}$, where $r$ is the radial coordinate, at a late time ($t = 1000$). There are no fitting parameters. The results in all the panels are nondimensional (\ref{eq:nondim}).}
\label{fig:finite_soluble_verification}
\end{figure}

\subsection{Connections to experiments}\label{subsec:experiment_link}

Here, we discuss how the theory outlined in this text is connected to experimental configurations (for the broader discussion of surfactant parameters, see \S \ref{subsubsec:surf_params_values}). Just for this subsection, we discuss dimensional variables (see \ref{eq:nondim}). The experimental results we discuss are presented in \cite{Eshima26_letter}, where surfactant-laden drops of mean drop radius $\hat{r}_d \approx 22$ \textmu m are deposited onto an air-water-air (Savart) sheet of thickness $\hat{h}_i\approx 20-30$ \textmu m. 
The surfactants used were sodium alkyl sulphates of various carbon lengths to allow for the systematic variation of surfactant solubility $\hat{k}_a \hat{k}_d^{-1}$. The experimental results, where the surfactant front evolution was tracked, agreed well with the theory developed here and in \cite{Eshima26_letter}. In particular, the enhancement of front propagation due to solubility was seen clearly in the experiments.

\begin{figure}
\begin{center}
\includegraphics[width=\textwidth]{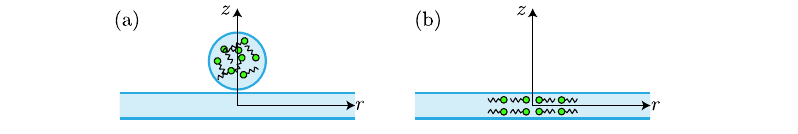}
\end{center}
\caption{A comparison of initial conditions between experimental and theoretical setups of surfactant deposition. (a) Initial condition of surfactant deposition due to the deposition of a surfactant-laden drop in experiment. (b) Initial condition of surfactant deposition in the theory (\ref{eq:ic}). The cross-film direction is given by $z$ and the radial direction is given by $r$.}
\label{fig:experiment_setup}
\end{figure}

Having further developed the theory in this paper, we can now note down some observations about comparing the experimental configuration to the theory. 
First, there are subtleties in the initial conditions. The time measurements are started when a surfactant-laden drop coalesces into the sheet (figure \ref{fig:experiment_setup}(a)). Then, the initial condition is different to that of (\ref{eq:ic}) where there is some surfactant distribution on a (flat) sheet otherwise at rest ((figure \ref{fig:experiment_setup}(b)), since the drop coalescence imposes some initial velocity. However, since we are interested in the late-time dynamics, the transient dynamics is believed to be not significant.
Nevertheless, in order for the theory to be applicable, some model for the effective initial distribution $\hat{\Gamma}_i$ and $\hat{c}_i$ must be identified, though as shown below, for the surfactant front location $\hat{r}_f$, such details are not needed. The initial distribution of surfactants firstly sets $\hat{n}_m$, which sets the horizontal characteristic length $\mathcal{L}_c$; see (\ref{eq:char_L}). Additionally, the initial distribution of surfactants sets the shape of the thickness profile $h$ in region I; see (\ref{eq:f_regI_finite_soluble}). The minimum thickness $h_{\text{min}}=\eta_f^{-2}t^{-1}$ and surfactant front location $r_f = \eta_f t^{1/2}$ in dimensional variables are given by (recall equations (\ref{eq:char_surf}, \ref{eq:def_Lambda_d}, \ref{eq:def_surf_deficit}, \ref{eq:char_L}, \ref{eq:char_T}))
\begin{subeqnarray}
   \hat{h}_{\text{min}} &=& \hat{h}_i\eta_f^{-2} \left(\frac{\hat{t}}{\mathcal{T}_c}\right)^{-1}  = \eta_f^{-2} \left(\frac{\rho \hat{h}_i^3 N}{\left(-\frac{d\hat{\sigma}}{d\hat{\Gamma}}|_{\hat{\Gamma}=0}\right) \pi \hat{n}_m^2}\frac{1+2\Lambda_d}{2\Lambda_d}\right)^{1/2}\hat{t}^{-1}\\
   \hat{r}_f &=& \eta_f \mathcal{L}_c\left(\frac{\hat{t}}{\mathcal{T}_c}\right)^{1/2} = \eta_f\left(\frac{\left(-\frac{d\hat{\sigma}}{d\hat{\Gamma}}|_{\hat{\Gamma}=0}\right)N}{\pi\rho \hat{h}_i}\frac{2\Lambda_d}{1+2\Lambda_d}\right)^{1/4}\hat{t}^{1/2}.
\end{subeqnarray}
Then, by (\ref{eq:eta_f_soluble}) and recalling definitions of $\mathcal{M}$ and $\Rey$ in (\ref{eq:def_M_Re}), we have
\allowdisplaybreaks[1]
\begin{subequations}
\begin{align}
   \hat{h}_{\text{min}} &=  \eta_f\left(\frac{\left(-\frac{d\hat{\sigma}}{d\hat{\Gamma}}|_{\hat{\Gamma}=0}\right)N}{\pi \hat{h}_i^2 \Sigma}, \sqrt{\frac{\left(-\frac{d\hat{\sigma}}{d\hat{\Gamma}}|_{\hat{\Gamma}=0}\right)\rho N}{\pi \hat{h}_i \mu^2}} , \infty\right)^{-2} \nonumber\\*
   & \times \left(\frac{\rho \hat{h}_i^3 N}{\left(-\frac{d\hat{\sigma}}{d\hat{\Gamma}}|_{\hat{\Gamma}=0}\right) \pi \hat{n}_m^2}\right)^{1/2}\hat{t}^{-1}\label{eq:hmin_dimensional_full}\\
   \hat{r}_f &=  \eta_f\left(\frac{\left(-\frac{d\hat{\sigma}}{d\hat{\Gamma}}|_{\hat{\Gamma}=0}\right)N}{\pi \hat{h}_i^2 \Sigma}, \sqrt{\frac{\left(-\frac{d\hat{\sigma}}{d\hat{\Gamma}}|_{\hat{\Gamma}=0}\right)\rho N}{\pi \hat{h}_i \mu^2}} , \infty\right) \nonumber\\*
   & \times \left(\frac{\left(-\frac{d\hat{\sigma}}{d\hat{\Gamma}}|_{\hat{\Gamma}=0}\right)N}{\pi\rho \hat{h}_i}\right)^{1/4}\hat{t}^{1/2}.\label{eq:rf_dimensional_full}
\end{align}
\end{subequations}
It can be seen from (\ref{eq:rf_dimensional_full}) that $\hat{r}_f$ does not require any information about the initial distribution, other than the total amount of surfactant $2N$. On the other hand, a value for $\hat{n}_m$ is additionally needed for (\ref{eq:hmin_dimensional_full}). 

A potential model that can be taken is as follows. Suppose that a drop of radius $\hat{r}_d$ is produced from a surfactant solution of bulk concentration $\hat{c}_0$. The drop produced also has concentration $\hat{c}_0$ with the total amount of surfactant $2N = (4/3) \pi \hat{r}_d^3\hat{c}_0$. Then, setting aside the details of the coalescence, a reasonable model to take for the initial condition (\ref{eq:ic}) in the theory is to let the initial condition be that of (\ref{eq:ic}), with $\hat{\Gamma}_i(\hat{r}) = 0$ and
\begin{equation}
    \hat{c}_i(\hat{r}) = \begin{cases} \hat{c}_0 & 0 \leq \hat{r} \leq \sqrt{\dfrac{4\hat{r}_d^3}{3\hat{h}_i}} \\[10pt] 0 & \hat{r} > \sqrt{\dfrac{4\hat{r}_d^3}{3\hat{h}_i}}, \end{cases}\label{eq:c_i_pancake}
\end{equation}
with the edge of the surfactant-containing region chosen by conservation of surfactant with total amount of surfactant equal to $(4/3) \pi \hat{r}_d^3\hat{c}_0$. For numerical simulations, it may be best to smooth the discontinuity for (\ref{eq:c_i_pancake}), but for illustrative purposes, the simple form (\ref{eq:c_i_pancake}) is chosen.  Then, according to the convention of (\ref{eq:def_surf_deficit}), since $\hat{n}_m = \hat{h}_i \hat{c}_0/2$, it follows that
\begin{equation}
    \Delta \Sigma_c = \left.-\frac{d\hat{\sigma}}{d\hat{\Gamma}}\right|_{\hat{\Gamma}=0}\frac{\Lambda_d\hat{h}_i\hat{c}_0}{1+2\Lambda_d} = \left.-\frac{d\hat{\sigma}}{d\hat{\Gamma}}\right|_{\hat{\Gamma}=0} \frac{\hat{k}_a\hat{k}_d^{-1}\hat{c}_0}{1+2\Lambda_d} .\label{eq:dsig_c_pancake}
\end{equation}

The most important detail to recognize when comparing theory to experiments is the surface tension deficit. In experiments, the surface tension deficit that is typically measured is that of the surfactant solution from which the drop is produced. For example, consider a surfactant solution of bulk concentration $\hat{c}_0$. Then, at equilibrium, the surface surfactant concentration $\hat{\Gamma}_0$ satisfies $\hat{k}_d \hat{\Gamma}_0 = \hat{k}_a \hat{c}_0$. Thus, the measured equilibrium surface tension deficit satisfies
\begin{equation}
    \Delta \Sigma_0 = -\left.\frac{d\hat{\sigma}}{d\hat{\Gamma}}\right|_{\hat{\Gamma}=0} \hat{k}_a \hat{k}_d^{-1} \hat{c}_0.\label{eq:dsig_0}
\end{equation}
In this paper, we define $\Delta \Sigma_c$ from the initial surfactant distributions $\hat{\Gamma}_i$ and $\hat{c}_i$ (\ref{eq:def_surf_deficit}). Assuming the model with $\hat{\Gamma}_i(\hat{r})=0$ and $\hat{c}_i(\hat{r})$ given by (\ref{eq:c_i_pancake}), we then have from (\ref{eq:dsig_c_pancake}) and (\ref{eq:dsig_0}) that
\begin{equation}
    \Delta \Sigma_c = \frac{\Delta \Sigma_0}{1+2\Lambda_d}
\end{equation}
and hence (\ref{eq:effective_st_transform}) gives
\begin{equation}
    \Delta \Sigma_{c,\text{eff}}=\frac{\Delta \Sigma_0}{2\Lambda_d},
\end{equation}
which is the expression identified by \cite{Eshima26_letter}.

Here, we discuss the relationships to, and consistency with, other nondimensionalisation conventions.
The choice of $\Delta \Sigma_c$ for the current manuscript makes sense since a characteristic surface tension must be defined from the initial surfactant distributions $\hat{\Gamma}_i$ and $\hat{c}_i$. In an experiment, by contrast (as above), the surface tension deficit of the surfactant solution, $\Delta \Sigma_0$, is a natural choice for the nondimensionalisation. Another option is to nondimensionalise using $\Delta \Sigma_{c,\text{eff}}$, which is effectively the surface tension deficit obtained by putting all the surfactants on the surface (\ref{eq:effective_st_transform}). However, the infinitely soluble limit is then not well defined and it is less clear that the infinitely soluble surfactant case can be recovered as $\Lambda_d \rightarrow 0$ as is done in \S \ref{sec:limiting_behaviours}. 

\section{Self-consistency for $0<\Lambda_d \ll 1$ via intermediate late-time solutions}\label{sec:limiting_behaviours}

\begin{figure}
\begin{center}
\includegraphics[width=\textwidth]{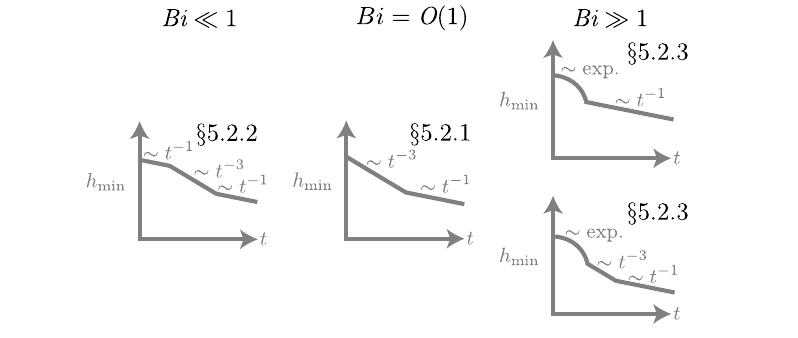}
\end{center}
\caption{A summary of the different thinning behaviours identified for the highly soluble limit $0<\Lambda_d \ll 1$. All cases ultimately satisfy minimum thickness $h_{\text{min}}=\eta_f(\mathcal{M}, \Rey, \Lambda_d)^{-2}t^{-1}$ for large enough times $t$ as identified in \S \ref{sec:finitely_soluble_surfactants}, but there are intermediate thinning behaviours. All schematic plots of $h_{\text{min}}$ against $t$ are log-log plots. Section numbers reference where the details are discussed, \S \S \ref{subsubsec:Bi1}, \ref{subsubsec:Bi0p01}, \ref{subsubsec:Bi100}.}
\label{fig:regime_Ld_small}
\end{figure}

In this section, we show how the solution we have identified for $0<\Lambda_d \ll 1$ in \S \ref{sec:finitely_soluble_surfactants} recovers the case of infinitely soluble surfactants as an intermediate late-time behaviour. For $0<\Lambda_d \ll 1$, the thinning dynamics follows that of infinitely soluble surfactants at intermediate late times prior to eventually crossing over to the similarity solution $h_{\text{min}}=\eta_f(\mathcal{M}, \Rey, \Lambda_d)^{-2}t^{-1}$ as expected through (\ref{eq:soluble_regIII}, \ref{eq:soluble_regIII_BC}). Scalings for the cross-over times between the behaviours are given. A summary of the dynamics identified is given in figure \ref{fig:regime_Ld_small}. Recall that the equations are non-dimensional; see (\ref{eq:nondim}). 

\subsection{Case $\Lambda_d = 0$}\label{subsec:Ld0}
We first consider the case $\Lambda_d = 0$. This case is not physically well-defined in our problem setup since the nondimensionalisation of the surface surfactant concentration $\hat{\Gamma}_c$ (\ref{eq:char_surf}) relies on $\Lambda_d$ being nonzero. Nonetheless, it is necessary to discuss the $\Lambda_d = 0$ case in order to understand cases with $0<\Lambda_d \ll 1$. The momentum and mass equations are given by (\ref{eq:tf_mom}, \ref{eq:tf_mass}) and the governing surface and bulk transport equations are given by 
\begin{subeqnarray}
    \frac{\partial \Gamma}{\partial t}&=&-\frac{1}{r}\frac{\partial }{\partial r}\left(r u \Gamma\right) + \textit{Bi}(c-\Gamma),\slabel{eq:surface_surf_Ld0}\\
    \frac{\partial c}{\partial t}+u\frac{\partial c}{\partial r}&=&0.\slabel{eq:bulk_surf_Ld0}
\label{eq:surf_transport_semi_infinitely_soluble}
\end{subeqnarray}

As in the infinitely soluble case \S \ref{subsec:inf_soluble}, equation (\ref{eq:bulk_surf_Ld0}) implies that $c$ remains $\textit{O}(1)$ behind the surfactant front and hence the thinning is much stronger than the insoluble case. However, the presence of a non-infinite $\textit{Bi}$ means that $\Gamma$ does not exactly equal $c$ and this leads to a weaker thinning than the infinitely soluble case. 

\begin{figure}
\begin{center}
\includegraphics[width=\textwidth]{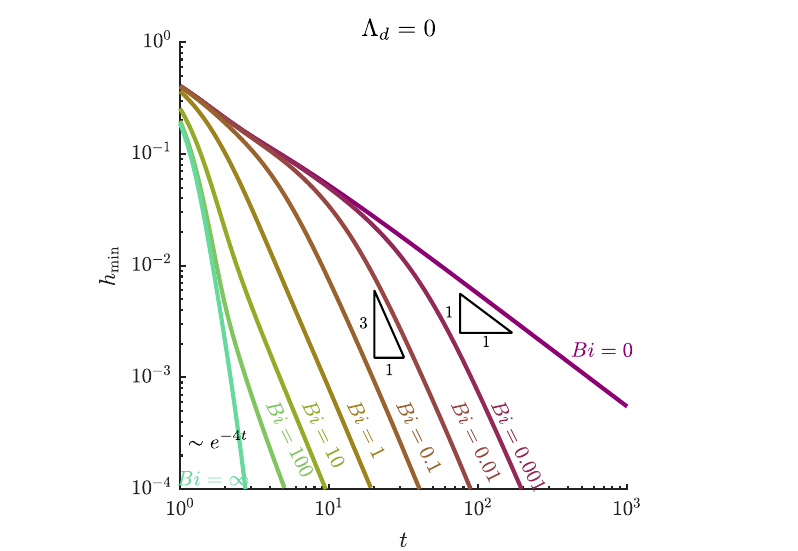}
\end{center}
\caption{Time evolution of the minimum thickness $h_{\text{min}}$ with $\Gamma_i = c_i = e^{-r^2},\mathcal{M}=1,\Rey=10$, and $\Lambda_d=0$. The curves are the numerical solutions of the thin-film equations (\ref{eq:tf_mom}, \ref{eq:tf_mass}, \ref{eq:surface_surf_Ld0}, \ref{eq:bulk_surf_Ld0}), coloured sequentially from magenta to blue for $\textit{Bi}=0,0.001,0.01,0.1,1,10,100,\infty$. The $\textit{Bi}=0$ case corresponds to the insoluble limit with $t^{-1}$ thinning as shown in figure \ref{fig:insoluble_inf_soluble_verification}(a) and $\textit{Bi}=\infty$ corresponds to infinitely soluble limit with exponential thinning as shown in figure \ref{fig:insoluble_inf_soluble_verification}(b). All $0<\textit{Bi}<\infty$ cases show $t^{-3}$ late-time thinning. The results shown are nondimensional (\ref{eq:nondim}).}
\label{fig:regime_Ld0}
\end{figure}

The time evolution of $h_{\text{min}}$ for $\textit{Bi}=0,0.001,0.01, 0.1, 1, 10, 100, \infty$ is shown in  figure \ref{fig:regime_Ld0}, where the different cases are coloured sequentially from magenta to blue. First, $\textit{Bi}=0$ corresponds to the insoluble case and is the same curve as shown in figure \ref{fig:insoluble_inf_soluble_verification}(a), which shows $t^{-1}$ thinning. Also, $\textit{Bi}=\infty$ corresponds to the infinitely soluble case and is the same curve as shown in figure \ref{fig:insoluble_inf_soluble_verification}(b), which shows exponential thinning. For $0<\textit{Bi}<\infty$, there is $t^{-3}$ thinning. Scaling arguments in Appendix \ref{app:scaling_Ld0} show that $h_{\text{min}}\sim \textit{Bi}^{-1}t^{-3}$ for $0<\textit{Bi}<\infty$. The singularity of the prefactor $\textit{Bi}^{-1}$ at $\textit{Bi}=0$ and $\textit{Bi}=\infty$ makes sense, given the respective changes in the scaling laws.

\subsection{Case $0<\Lambda_d \ll 1$}
\subsubsection{Case $\textit{Bi}=\textit{O}(1)$}\label{subsubsec:Bi1}

When $\textit{Bi}=\textit{O}(1)$ and $0<\Lambda_d \ll 1$, the intermediate late time behaviour is  $t^{-3}$ thinning, as is the case for $\Lambda_d = 0$ (\S \ref{subsec:Ld0}). However, since $\Lambda_d \neq 0$, beyond sufficiently late times, $t^{-1}$ thinning is once again recovered. The cross-over time $t_c$ is estimated by comparing $h_{\text{min}}\sim t^{-3}$ ($\textit{Bi}\neq 0$, $\Lambda_d = 0$)  to $h_{\text{min}} =2^{-3/2}\Lambda_d^{1/2}t^{-1}$ (\ref{eq:eta_f_Ld_small_limit}) to obtain $t_c=\textit{O}( \Lambda_d^{-1/4})$.

\begin{figure}
\begin{center}
\includegraphics[width=\textwidth]{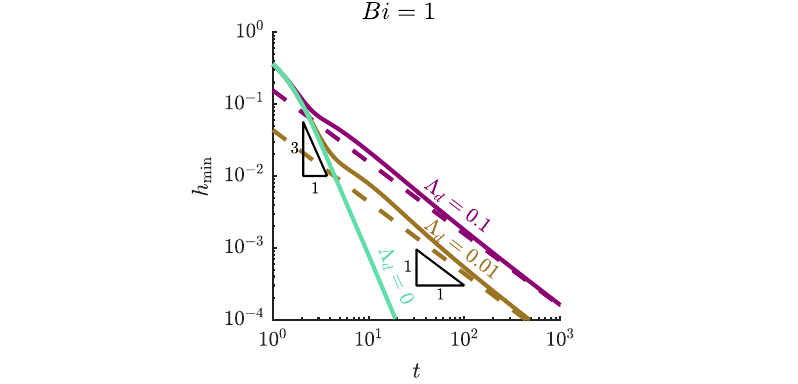}
\end{center}
\caption{Time evolution of the minimum thickness $h_{\text{min}}$ with $\Gamma_i = c_i = e^{-r^2},\mathcal{M}=1,\Rey=10,\textit{Bi}=1$. The solid curves show the numerical solutions of the thin-film equations (\ref{eq:tf_mom}, \ref{eq:tf_mass}, \ref{eq:tf_surface_surf}, \ref{eq:tf_bulk_surf}) for $\Lambda_d = 0.1,0.01,0$ (magenta, gold-brown, blue). Dashed lines show the similarity solution for $\Lambda_d \neq 0$ calculated via (\ref{eq:soluble_regIII},\ref{eq:soluble_regIII_BC}). There are no fitting parameters. The results shown are nondimensional (\ref{eq:nondim}).}
\label{fig:regime_Bi1}
\end{figure}

The time evolution of $h_{\text{min}}$ is shown in figure \ref{fig:regime_Bi1} for $\textit{Bi}=1$ and $\Lambda_d = 0.1, 0.01, 0$ (magenta, gold-brown, blue). Prior to $t_c$, the $\Lambda_d = 0.1, 0.01$ cases satisfy the same $t^{-3}$ thinning as for $\Lambda_d = 0$. Beyond $t_c$, $t^{-1}$ thinning is seen and agrees with the similarity solution (dashed lines) calculated via (\ref{eq:soluble_regIII},\ref{eq:soluble_regIII_BC}).

\subsubsection{Case $\textit{Bi}\ll 1$}\label{subsubsec:Bi0p01}
When $\textit{Bi}\ll 1$ and $0< \Lambda_d \ll 1$, the first intermediate late-time behaviour is $t^{-1}$, as for $\textit{Bi}=0$. However, since $\textit{Bi}\neq 0$, there is eventually $t^{-3}$ thinning, as for $\Lambda_d = 0$. The first cross-over time $t_{c1}$ is estimated by comparing $h_{\text{min}}\sim t^{-1}$ ($\textit{Bi}=0$) to $h_{\text{min}}\sim \textit{Bi}^{-1}t^{-3}$ ($\textit{Bi}\neq 0,\Lambda_d=0$), which leads to $t_{c1} = \textit{O}(\textit{Bi}^{-1/2})$. Finally, for sufficiently large times, $t^{-1}$ thinning behaviour is once again recovered. The second cross-over time $t_{c2}$ is estimated by comparing $h_{\text{min}}\sim \textit{Bi}^{-1}t^{-3}$ ($\textit{Bi}\neq 0,\Lambda_d=0$) to $h_{\text{min}} = 2^{-3/2}\Lambda_d^{1/2}t^{-1}$ (\ref{eq:eta_f_Ld_small_limit}), which leads to $t_{c2} = \textit{O}(\textit{Bi}^{-1/2}\Lambda_d^{-1/4})$.

\begin{figure}
\begin{center}
\includegraphics[width=\textwidth]{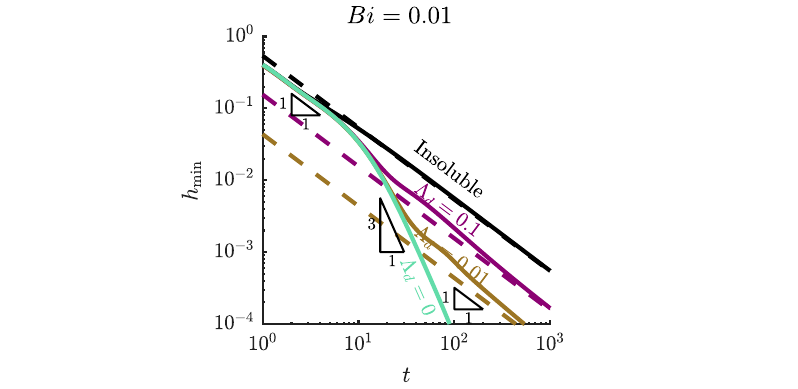}
\end{center}
\caption{Time evolution of the minimum thickness $h_{\text{min}}$ with $\Gamma_i = c_i = e^{-r^2},\mathcal{M}=1,\Rey=10$, and $\textit{Bi}=0.01$. The coloured solid curves show the numerical solutions of the thin-film equations (\ref{eq:tf_mom}, \ref{eq:tf_mass}, \ref{eq:tf_surface_surf}, \ref{eq:tf_bulk_surf}) for $\Lambda_d = 0.1,0.01,0$ (magenta, gold-brown, blue). The black curve shows the insoluble surfactant deposition limit, which is the same curve as figure \ref{fig:insoluble_inf_soluble_verification}(a). Dashed lines show the similarity solution for $\Lambda_d \neq 0$ calculated via (\ref{eq:soluble_regIII},\ref{eq:soluble_regIII_BC}). There are no fitting parameters. The results shown are nondimensional (\ref{eq:nondim}).}
\label{fig:regime_Bi0.01}
\end{figure}

The time evolution of $h_{\text{min}}$ is shown in figure \ref{fig:regime_Bi0.01} for $\textit{Bi}=0.01$ and $\Lambda_d = 0.1, 0.01, 0$ (magenta, gold-brown, blue). The black solid curve denotes the insoluble surfactant limit, which is the same curve as figure \ref{fig:insoluble_inf_soluble_verification}(a). 
Prior to $t_{c1}$, the $\Lambda_d = 0.1,0.01$ cases satisfy the same $t^{-1}$ thinning as for insoluble surfactants. Beyond $t_{c1}$ but prior to $t_{c2}$, $t^{-3}$ thinning is seen, as for $\Lambda_d = 0$. Finally, beyond $t_{c2}$, the $t^{-1}$ thinning is seen and agrees with the similarity solution (dashed lines) calculated via (\ref{eq:soluble_regIII},\ref{eq:soluble_regIII_BC}). 

\subsubsection{Case $\textit{Bi}\gg 1$}\label{subsubsec:Bi100}

When $\textit{Bi}\gg 1$ and $0< \Lambda_d \ll 1$, the first intermediate late-time behaviour is exponential thinning, as for the infinitely soluble limit, which leads to the first cross-over time $t_{c1}$ being practically $\textit{O}(1)$. There are then two possible scenarios. By comparing $h_{\text{min}}\sim \textit{Bi}^{-1}t^{-3}$ ($\textit{Bi}\neq 0,\Lambda_d=0$) and $h_{\text{min}}=2^{-3/2}\Lambda_d^{1/2}t^{-1}$ (\ref{eq:eta_f_Ld_small_limit}), if $\Lambda_d \gg \textit{Bi}^{-2}$, then the final $t^{-1}$ thinning is achieved beyond $t_{c1}$ (no $t^{-3}$ thinning). If $\Lambda_d \ll \textit{Bi}^{-2}$, then there is $t^{-3}$ thinning before the final $t^{-1}$ thinning with the second cross-over time $t_{c2} = \textit{O}(\textit{Bi}^{-1/2}\Lambda_d^{-1/4})$.

\begin{figure}
\begin{center}
\includegraphics[width=\textwidth]{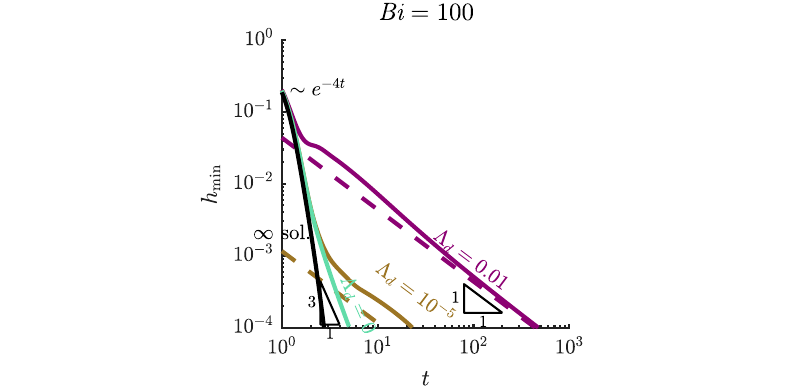}
\end{center}
\caption{Time evolution of the minimum thickness $h_{\text{min}}$ with $\Gamma_i = c_i = e^{-r^2},\mathcal{M}=1,\Rey=10$, and $\textit{Bi}=100$. The solid curves show the numerical solutions of the thin-film equations (\ref{eq:tf_mom}, \ref{eq:tf_mass}, \ref{eq:tf_surface_surf}, \ref{eq:tf_bulk_surf}) for $\Lambda_d = 0.01,10^{-5},0$ (magenta, gold-brown, blue). Dashed lines show the similarity solution for $\Lambda_d = 0.01, 10^{-5}$ calculated via (\ref{eq:soluble_regIII}, \ref{eq:soluble_regIII_BC}). The black curve shows the time evolution of the minimum thickness for the infinitely soluble limit (\ref{eq:tf_mom}, \ref{eq:tf_mass}, \ref{eq:inf_soluble_transport}).
There are no fitting parameters. The results shown are nondimensional (\ref{eq:nondim}).}
\label{fig:regime_Bi100}
\end{figure}

Time evolution of $h_{\text{min}}$ is shown in figure \ref{fig:regime_Bi100} for $\textit{Bi}=100$ and $\Lambda_d = 0.01, 10^{-5}, 0$ (magenta, gold-brown, blue). The black curve shows the infinitely soluble limit, which is the same as that shown in figure \ref{fig:insoluble_inf_soluble_verification}(b). Prior to $t_{c1}$, the $\Lambda_d = 0.01, 10^{-5}$ cases show exponential thinning as for infinitely soluble surfactants. Beyond $t_{c1}$, the $\Lambda_d = 0.01$ case reaches the final $t^{-1}$ thinning without $t^{-3}$ thinning, which agrees with the similarity solution (dashed lines) calculated via (\ref{eq:soluble_regIII},\ref{eq:soluble_regIII_BC}). Beyond $t_{c1}$ but prior to $t_{c2}$, the $\Lambda_d=10^{-5}$ case shows $t^{-3}$ thinning, as for $\Lambda_d = 0$. Beyond $t_{c2}$, the $\Lambda_d = 10^{-5}$ case shows $t^{-1}$ thinning. The $\Lambda_d = 10^{-5}$ case is slow to asymptote to the similarity solution (dashed lines) calculated via (\ref{eq:soluble_regIII},\ref{eq:soluble_regIII_BC}). Agreement is expected to improve at longer times, which we could not demonstrate here due to numerical limitations.  

\section{Inclusion of cross-film diffusion}\label{sec:cross-film_diffusion}

In this section, we show how cross-film diffusion can be included by considering a nonzero $\textit{Pe}_{\perp}$. The thin-film equations we consider for $u(r,t), h(r,t), \Gamma(r,t), c(r,z,t)$ are given by (\ref{eq:tf_mom}, \ref{eq:tf_mass}, \ref{eq:tf_surface_surf}, \ref{eq:tf_bulk_surf_2D}) respectively. Recall that the equations are non-dimensional; see (\ref{eq:nondim}). It will be shown that the late-time similarity solution is independent of $\textit{Pe}_{\perp}$.

In order to solve the thin-film equations numerically, we first introduce the variable 
\begin{equation}
    s_z:=\frac{z}{h(r,t)}.
\end{equation}
For extensional flow where the cross-film velocity is linear with respect to $z$, $s_z$ is the cross-film Lagrangian coordinate. Then, transforming equation (\ref{eq:tf_bulk_surf_2D}) from $(r,z,t)\rightarrow (r,s_z,t)$,
\begin{equation}
    \frac{\partial c}{\partial t}  -\frac{\partial h}{\partial t}\frac{s_z}{h}\frac{\partial c}{\partial s_z}+ u\left(\frac{\partial c}{\partial r}-\frac{\partial h}{\partial r}\frac{s_z}{h}\frac{\partial c}{\partial s_z}\right) -s_z \left(\frac{u}{r}+\frac{\partial u}{\partial r}\right) \frac{\partial c}{\partial s_z} = \textit{Pe}_{\perp}^{-1}h^{-2}\frac{\partial^2 c}{\partial s_z^2}
\end{equation}
and hence by conservation of mass (\ref{eq:tf_mass}),
\begin{equation}
    \frac{\partial c}{\partial t}+u\frac{\partial c}{\partial r}=\textit{Pe}_{\perp}^{-1}h^{-2}\frac{\partial^2 c}{\partial s_z^2}.\label{eq:c_evol_eq_sz}
\end{equation}
Similarly, the cross-film boundary conditions (\ref{eq:c_cross-film_bc_z0},\ref{eq:c_cross-film_bc_interface}) become:
\begin{equation}
    \left.\frac{\partial c}{\partial s_z}\right|_{s_z=0}=0,~\left.\frac{\partial c}{\partial s_z}\right|_{s_z=1/2}= -\textit{Pe}_{\perp}\textit{Bi}\Lambda_d h(c|_{s_z=1/2}-\Gamma).
\end{equation}

\begin{figure}
\begin{center}
\includegraphics[width=\textwidth]{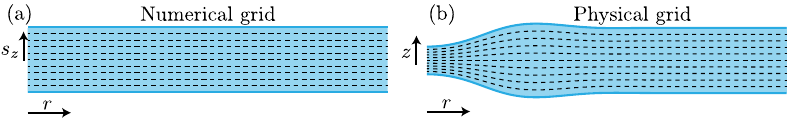}
\end{center}
\caption{Cross-film Lagrangian coordinate transformation used for solving the numerical solution of the thin-film equations (\ref{eq:tf_mom}, \ref{eq:tf_mass}, \ref{eq:tf_surface_surf}, \ref{eq:tf_bulk_surf_2D}). (a) Numerical grid given by $s_z = z/h(r,t)$ where $h$ is the thickness profile and $z$ is the cross-film coordinate. (b) Physical grid given by $z$. For both (a,b), the horizontal coordinate is the radial coordinate $r$ and the dashed curves show curves of constant $s_z$.}
\label{fig:2D_grid_transformation}
\end{figure}

The introduction of a cross-film Lagrangian coordinate to eliminate the cross-film advection term in (\ref{eq:tf_bulk_surf_2D}) is similar to the method introduced by \cite{Ranz79} to account for the effect of lamella stretching to liquid mixing. In our work, the transformation $(r,z,t)\rightarrow (r,s_z,t)$ is also numerically convenient since the $c(r,s_z,t)$ field is computed on a fixed rectangular grid $(r,s_z)\in [0,\infty]\times [0,1/2]$, which is displayed in figure \ref{fig:2D_grid_transformation}(a), instead of a computational domain with a moving boundary, as in figure \ref{fig:2D_grid_transformation}(b). Once the PDEs are solved in $(r,s_z,t)$ coordinates, the solution can be translated to $(r,z,t)$ coordinates via $z = s_z h$.

\begin{figure}
\begin{center}
\includegraphics[width=\textwidth]{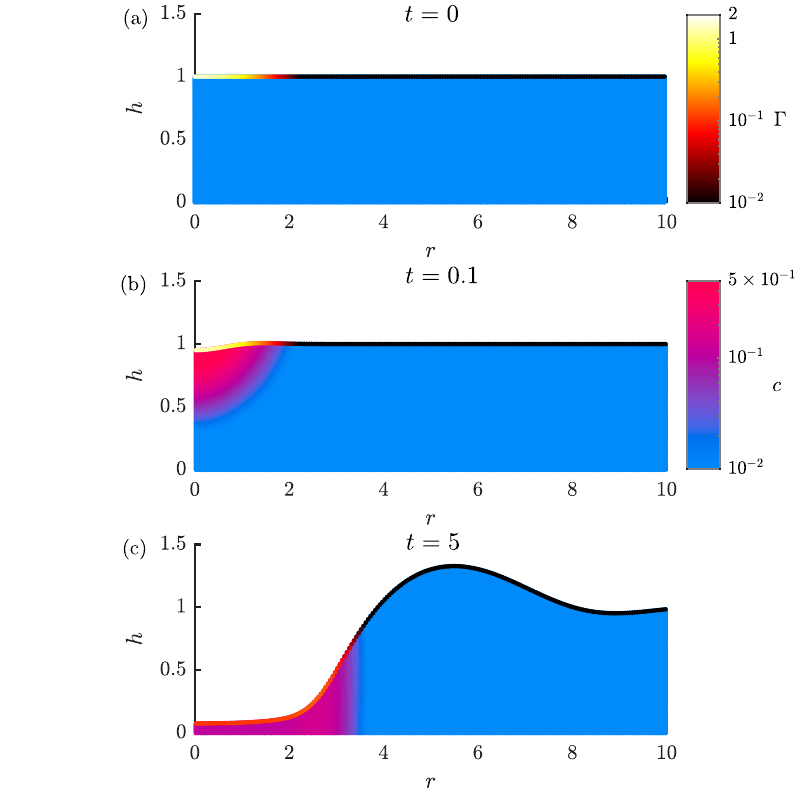}
\end{center}
\caption{Sample evolution for finitely soluble surfactant deposition with cross-film diffusion for $\Gamma_i = ((2\Lambda_d+1)/(2\Lambda_d))e^{-r^2}, c_i=0, \mathcal{M} = 1, \Rey = 10, \textit{Bi}=1, \Lambda_d = 1$, and $\textit{Pe}_{\perp}=10$, which show numerical solutions to the thin-film equations (\ref{eq:tf_mom}, \ref{eq:tf_mass}, \ref{eq:tf_surface_surf}, \ref{eq:tf_bulk_surf_2D}). The thickness $h$ is plotted against the radial coordinate $r$. The colours on the surface show the surface surfactant concentration $\Gamma$ (black to yellow, where values below $10^{-2}$ are set to black) and the colours in the bulk show the bulk surfactant concentration $c$ (blue to pink, where values below $10^{-2}$ are set to blue). (a) $t=0$. (b) $t = 0.1$. (c) $t = 5$. The results in all the panels are nondimensional (\ref{eq:nondim}).}
\label{fig:2D_sample_prop}
\end{figure}

Sample dynamics of the sheet are given in figure $\ref{fig:2D_sample_prop}$ for $\textit{Pe}_{\perp}=10$. In order to emphasise the cross-film inhomogeneity of $c$ at early times, we consider the initial condition where all the surfactants are on the surface, that is, $\Gamma_i = [(2\Lambda_d+1)/(2\Lambda_d)]e^{-r^2}$ and $c_i = 0$ (figure $\ref{fig:2D_sample_prop}$(a)); recall that the prefactor is set by the convention for nondimensionalisation. The colours show the surface and bulk surfactant concentrations. At $t = 0.1$, there is indeed cross-film inhomogeneity of $c$ (figure $\ref{fig:2D_sample_prop}$(b)). However, as time progresses, $c$ quickly becomes uniform in the cross-film direction (figure $\ref{fig:2D_sample_prop}$(c) shows $t = 5$), since diffusion homogenises distributions over time. Additionally, the film thins over time, which further enhances cross-film diffusion. The enhancement of diffusion can be seen in equation (\ref{eq:c_evol_eq_sz}) by identifying $\textit{Pe}_{\perp}h^2$ as an ``effective" Péclet number.

\begin{figure}
\begin{center}
\includegraphics[width=\textwidth]{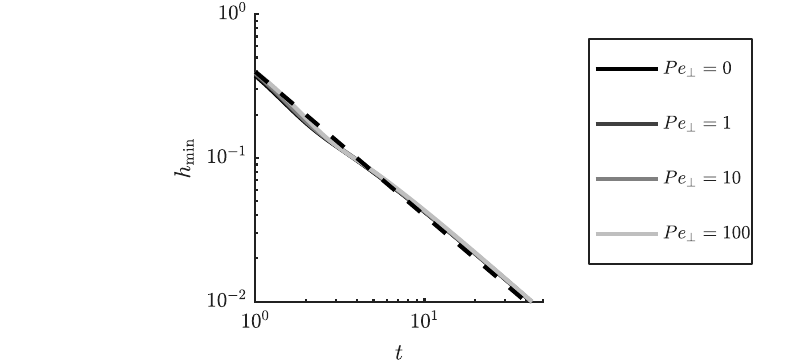}
\end{center}
\caption{Time evolution of the minimum thickness $h_{\text{min}}$ with $\Gamma_i = c_i = e^{-r^2},\mathcal{M}=1,\Rey=10,\textit{Bi}=1,\Lambda_d = 1$. The coloured solid curves show the numerical solutions of the thin-film equations (\ref{eq:tf_mom}, \ref{eq:tf_mass}, \ref{eq:tf_surface_surf}, \ref{eq:tf_bulk_surf_2D}) for $\textit{Pe}_{\perp} = 0,1,10,100$ (black to light gray). Dashed line shows the similarity solution predicted via (\ref{eq:soluble_regIII}, \ref{eq:soluble_regIII_BC}). The results shown are nondimensional (\ref{eq:nondim}).}
\label{fig:2D_hmin}
\end{figure}

Since $c$ becomes homogeneous in the cross-film direction at sufficiently long times, the late-time similarity solution is therefore analogous to that identified in \S \ref{sec:finitely_soluble_surfactants}. The evolution of the minimum thickness over time for $\textit{Pe}_{\perp}=0,1,10,100$ (black to light gray) is shown in figure \ref{fig:2D_hmin}. The dashed line shows the similarity solution $h_{\text{min}}=\eta_f(\mathcal{M},\Rey,\Lambda_d)^{-2}t^{-1}$ in (\ref{eq:soluble_regIII}, \ref{eq:soluble_regIII_BC}). It can be seen that there is once again good agreement with the late-time similarity solution. Furthermore, the late-time similarity solution is indeed independent of $\textit{Pe}_{\perp}$.   

\section{Conclusion}

In this paper, we considered the evolution of a thin air-liquid-air sheet due to finitely soluble surfactants, accounting for inertia, Marangoni stress, capillary stress, extensional stress, surface and bulk surfactant transport, surfactant adsorption-desorption, and cross-film diffusion. This paper numerically and theoretically extends the results in \cite{Eshima25_JFM,Eshima25_PRL, Eshima26_letter}.

In \S \ref{sec:problem_setup}, the full nondimensionalisation of the governing equations is given, taking particular care in defining the characteristic scales and nondimensional numbers. 
In \S \ref{sec:background_solns}, the two extreme limits of insoluble and infinitely soluble surfactants are discussed.
In \S \ref{subsec:deriv_simil_soln_fin_soluble}, the late-time similarity solution for finitely soluble surfactants is derived, where the analytical mapping (\ref{eq:eta_f_soluble}) to the equivalent insoluble surfactant deposition is shown. Thus, at late-times, the scalings for the minimum thickness $h_{\text{min}}\sim t^{-1}$ and front propagation $r_f \sim t^{1/2}$ do not change, but the prefactor changes and the effect of finite solubility is captured through a single nondimensional parameter $\Lambda_d$.
The mapping (\ref{eq:eta_f_soluble}) shows that the insoluble surfactant similarity solution can be derived from the finitely soluble similarity solution in the limit $\Lambda_d \rightarrow \infty$. 
In \S \ref{subsec:fin_sol_verif}, the numerical solutions of the thin-film equations are compared to the theoretical prediction, which shows good agreement.
In \S \ref{subsec:experiment_link}, we point out some details to consider when comparing the theoretical results to surfactant deposition experiments, with a focus on the definition of the characteristic surface tension deficit $\Delta \Sigma_c$.
In \S \ref{sec:limiting_behaviours}, we show that the similarity solution for finitely soluble surfactants is consistent with the similarity solution for infinitely soluble surfactants in the limit $\Lambda_d \rightarrow 0$ through intermediate late-time behaviour. 
In \S \ref{sec:cross-film_diffusion}, it is shown that accounting for the transient dynamics of cross-film diffusion does not change the late-time similarity solution.

There are many pieces of physics that can be included into the governing equations identified here to extend our results for thin films, such as background flows \citep{Burton07,Fontelos18}, van der Waals forces \citep{Vaynblat02,Wee22,Wee24}, and having multiple types of surfactants. Furthermore, it is hoped that the theoretical setup and analytical results in this paper could be used as a benchmark for direct numerical simulation codes (DNS) that are actively under development in the community, e.g., \cite{Muradoglu08,Shin18, Farsoiya24}, which solve the two-phase Navier-Stokes equations with surfactants. Our theory accounts for physical effects, such as inertia, capillarity, surface and bulk surfactant transport with deforming boundaries, which are challenging to implement numerically in DNS codes that do not assume a thin-film geometry.

By defining the non-dimensional parameters quantitatively, it is possible to connect theoretical and numerical solutions to physically relevant solutions. As discussed in \S \ref{subsubsec:surf_params_values}, surfactant properties can vary greatly between surfactants and it is therefore important to account for a wide range of surfactant parameters, as was investigated in this text using $\textit{Bi}, \Lambda_d, \textit{Pe}_{\perp}$.
Beyond thin films, it is hoped that this detailed approach can be applied to extend the quantitative understanding of surfactant effects on other physical problems (see e.g. \cite{Takagi11,Manikantan20}), such as bubble bursting, bubble breakup, bubble flotation, drop coalescence, filament fragmentation, and surface wave damping.

\begin{acknowledgments}
\textbf{Acknowledgments}.
We thank Tristan Aurégan and Emmanuel Villermaux for helpful discussions. Figures in this text used color maps by \cite{Crameri2021}.

This research was conducted under the Princeton SEAS Innovation Fund and NSF grant No. 2242512 to L.D.; the NSF Grant No. CBET-2246791 to H.A.S.; the Eli and Britt Harari Fellowship and the Wallace Memorial Fellowship at Princeton University to J.E. The simulations presented in this manuscript were performed on computational resources managed and supported by Princeton Research Computing, a consortium of groups including the Princeton Institute for Computational Science and Engineering (PICSciE) and Research Computing at Princeton University.
\end{acknowledgments} 

\textbf{Declaration of interests.} The authors report no conflict of interest.

\appendix
\section{Derivation of exponential factor for infinitely soluble surfactants}\label{app:inf_soluble_derivation}

For infinitely soluble surfactants with the thin-film equations given by (\ref{eq:tf_mom}, \ref{eq:tf_mass}, \ref{eq:inf_soluble_transport_surface}, \ref{eq:inf_soluble_transport_bulk}) for $u,h,\Gamma,c$, it is possible to obtain the thickness at the origin without solving the full similarity solution. In the case of a Gaussian initial condition $\Gamma_i = c_i = e^{-r^2}$ as considered in the example in \S \ref{subsec:inf_soluble}, the thickness at the origin is indeed the minimum thickness. We additionally assume in this derivation that $d^2 c_i/ds^2<0$ (i.e. a local maximum of surfactant at the origin). It can be checked numerically or by scaling arguments that the leading-order radial momentum balance (\ref{eq:tf_mom}) in the region of uniform surfactant concentration (i.e. near the origin, see figure \ref{fig:soluble_sample_dynamics}) is between inertia and Marangoni stress. Then,
\begin{equation}
    \frac{\partial u}{\partial t}+u\frac{\partial u}{\partial r}=-\frac{2}{h}\frac{\partial c}{\partial r}.
\end{equation}
Consider radial Lagrangian coordinates $(s,t)$ and let the Lagrangian time derivative be given by $D/Dt$. In particular, we have $\frac{\partial s}{\partial r}=\frac{rh}{s}$ (see Appendix \ref{app:horiz_Lag_coords}) and hence
\begin{equation}
    \left.\frac{Du}{Dt}\right|_{(s,t)}=-\frac{2r}{s}\frac{\partial c}{\partial s}.\label{eq:lord_mom_inf_sol}
\end{equation}
Also, note that (\ref{eq:inf_soluble_transport}) gives
\begin{equation}
    \left.\frac{Dc}{Dt}\right|_{(s,t)}=0
\end{equation}
and hence $c=c_i(s)$ for all times $t$. Then, upon differentiating (\ref{eq:lord_mom_inf_sol}) with respect to $t$ and using the expression $Dr/Dt = u$, we have
\begin{equation}
    \left.\frac{D^2u}{Dt^2}\right|_{(s,t)}=-\frac{2}{s}\frac{d c_i}{d s}u
\end{equation}
which can be integrated with respect to $t$ to give
\begin{equation}
    u = A_1(s) \exp\left(\left(-\frac{2}{s}\frac{dc_i}{ds}\right)^{1/2} t\right)+A_2(s) \exp\left(-\left(-\frac{2}{s}\frac{dc_i}{ds}\right)^{1/2} t\right)\label{eq:u_inf_soluble_regI}
\end{equation}
for some functions $A_1(s)$ and $A_2(s)$.

Now, the sheet is thin $h \ll 1$ near the origin at late times, $s\approx 0$ in this region (see e.g., (\ref{eq:horiz_lag_func})). Additionally at late times, the exponential with a negative exponent decreases (\ref{eq:u_inf_soluble_regI}). Noting that  $(dc_i/ds)/s|_{s=0}=d^2c_i/ds^2|_{s=0}$, we have near the origin 
\begin{equation}
    u \approx A\exp\left(\left(\left.-2\frac{d^2c_i}{ds^2}\right|_{s=0}\right)^{1/2} t\right).
\end{equation}
Then, since $Dr/Dt = u$, we have that close to the origin,
\begin{equation}
    u \approx \left(\left.-2\frac{d^2c_i}{ds^2}\right|_{s=0}\right)^{1/2} r.
\end{equation}
Indeed, the linear relation to $u$ with respect to $r$ can be seen in figure \ref{fig:insoluble_infsoluble_sample_dynamics}(f). Then,
\begin{equation}
    \left.\frac{\partial u}{\partial r}\right|_{r=0}=\left(\left.-2\frac{d^2c_i}{ds^2}\right|_{s=0}\right)^{1/2}.
\end{equation}
Also, from conservation of mass (\ref{eq:tf_mass}),
\begin{equation}
    \frac{d}{dt}h(0,t) = \left.-2\frac{\partial u}{\partial r}\right|_{r=0}h(0,t) = -2\left(\left.-2\frac{d^2c_i}{ds^2}\right|_{s=0}\right)^{1/2} h(0,t).
\end{equation}
Hence
\begin{equation}
    h(0,t) \sim \exp\left(-2\left(\left.-2\frac{d^2c_i}{ds^2}\right|_{s=0}\right)^{1/2}t\right).
\end{equation}
In the case of a Gaussian initial condition $c_i(s) = e^{-s^2}$, it then follows that $h(0,t) \sim e^{-4t}$, which indeed agrees with figure \ref{fig:insoluble_inf_soluble_verification}(b). 

\section{Radial Lagrangian coordinates to derive thickness profile in region I}\label{app:horiz_Lag_coords}

In this appendix, we derive the thickness profile in region I (\ref{eq:f_regI_finite_soluble}). The derivation is analogous to the insoluble surfactant case as described by \cite{Eshima25_JFM,Eshima25_PRL}. Let $s$ denote the radial Lagrangian coordinate, i.e. $s$ is the radial coordinate of a material element at initial time. Considering the Eulerian radial coordinate $r$ as a function of $(s,t)$ given by $r = r(s,t)$, we therefore have $s = r(s,0)$. By global conservation of mass, it follows that $s$ can also be written as a function of $(r,t)$, given explicitly by 
\begin{equation}
    s = \left(2\int_0^r R h(R,t) dR\right)^{1/2}.\label{eq:horiz_lag_func}
\end{equation}

Let the Lagrangian time derivative be given by $D/Dt$. Then, in Lagrangian coordinates $(s,t)$,  conservation of mass (\ref{eq:tf_mass}), surface surfactant concentration (\ref{eq:tf_surface_surf}), and bulk surfactant  concentration (\ref{eq:tf_bulk_surf}) become
\begin{subeqnarray}
    \left.\frac{Dh}{Dt}\right|_{(s,t)} &=& \left.\left(-\frac{hu}{r} - h\frac{\partial u}{\partial r}\right)\right|_{r(s,t),t}\\
    \left.\frac{D\Gamma}{Dt}\right|_{(s,t)} &=& \left.\left(-\frac{\Gamma u}{r} - \Gamma\frac{\partial u}{\partial r} + \textit{Bi}(c-\Gamma)\right)\right|_{r(s,t),t},\\
    \left.\frac{Dc}{Dt}\right|_{(s,t)} &=& \left.-\frac{2}{h}\textit{Bi}\Lambda_d (c-\Gamma)\right|_{r(s,t),t}.\label{eq:lag_coords}
\end{subeqnarray}
Then, it follows directly from (\ref{eq:lag_coords}) that
\begin{equation}
    \left.\frac{D}{Dt}\left(\frac{2\Lambda_d}{2\Lambda_d+1}\frac{\Gamma}{h}+\frac{1}{2\Lambda_d+1}c\right)\right|_{(s,t)}=0.
\end{equation}
Thus, it follows exactly from the initial condition (\ref{eq:ic}) that at all times
\begin{equation}
    \left.\left(\frac{2\Lambda_d}{2\Lambda_d+1}\frac{\Gamma}{h}+\frac{1}{2\Lambda_d+1}c\right)\right|_{(s,t)} = \frac{2\Lambda_d}{2\Lambda_d+1}\Gamma_i(s)+\frac{1}{2\Lambda_d+1}c_i(s).
\end{equation}
Substituting the late-time solutions (\ref{eq:sol_regI}) gives
\begin{equation}
    \frac{1}{f(\eta)} = \frac{2\Lambda_d}{2\Lambda_d +1}\Gamma_i(s(\eta)) + \frac{1}{2\Lambda_d+1}c_i(s(\eta))
\end{equation}
to leading order where $s = \left(2\int_0^{\eta} \eta'f(\eta') d \eta'\right)^{\frac{1}{2}}$ is the radial Lagrangian coordinate.

\section{Alternate initial conditions}\label{app:different_ic}

Throughout \S \S \ref{sec:finitely_soluble_surfactants},\ref{sec:limiting_behaviours}, we only consider examples that start at kinetic equilibrium $\Gamma_i = c_i$ since the details of the initial conditions do not change the late-time similarity solutions. In this appendix, we give an example where the system starts away from kinetic equilibrium.

\begin{figure}
\begin{center}
\includegraphics[width=\textwidth]{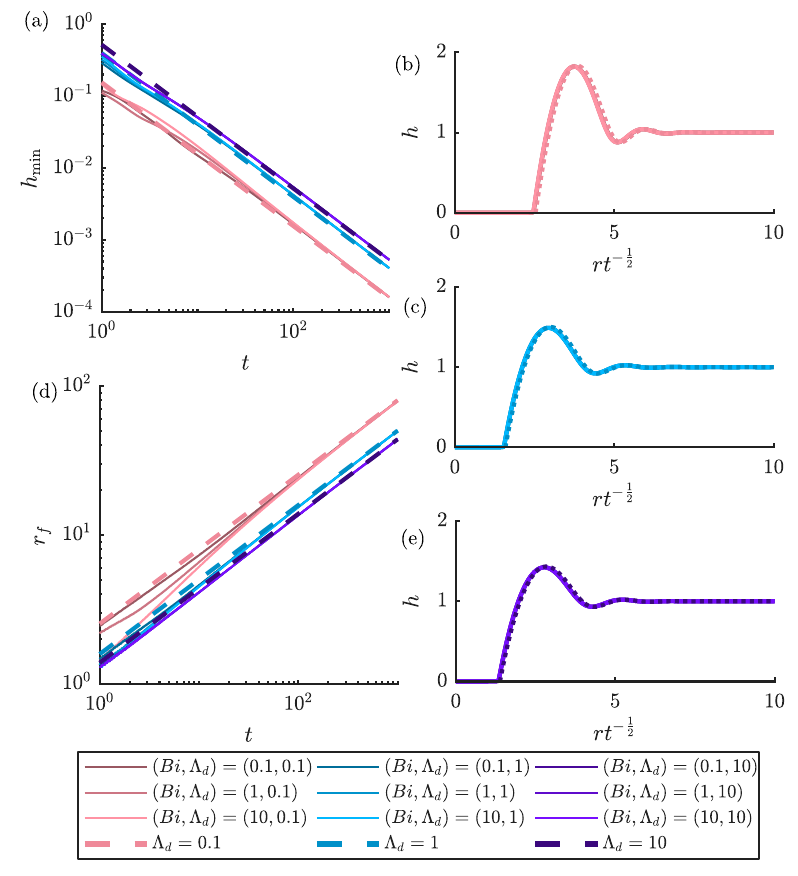}
\end{center}
\caption{Finitely soluble surfactant deposition with $\Gamma_i=\frac{2\Lambda_d+1}{2\Lambda_d}e^{-r^2}, c_i = 0,\mathcal{M}=1,\Rey=10$. The panel conventions are the same as figure \ref{fig:finite_soluble_verification}. Again, there are no fitting parameters. The results in all the panels are nondimensional (\ref{eq:nondim}).}
\label{fig:finite_soluble_verification_all_surface}
\end{figure}

The parameter sweep of $\textit{Bi},\Lambda_d$ for $\Gamma_i=\frac{2\Lambda_d+1}{2\Lambda_d}e^{-r^2}, c_i = 0$ is shown in figure \ref{fig:finite_soluble_verification_all_surface}. In other words, we consider the initial condition of a Gaussian deposition that starts with all the surfactants on the interface; recall that the prefactor is set by the nondimensionalisation convention (\ref{eq:nondim}). The panel conventions of figure \ref{fig:finite_soluble_verification_all_surface} are the same as figure \ref{fig:finite_soluble_verification}. As can be seen in the figure, there is good agreement with the late-time similarity solution once again without any fitting parameters. Thus the discussions in the main text hold regardless of the initial surfactant distribution.   

\section{High $\mathcal{M}, \Rey$ limit }\label{app:lim_highM_Re}

In this appendix, we analytically solve for the similarity solution (\ref{eq:soluble_regIII}, \ref{eq:soluble_regIII_BC}) for $(\mathcal{M},\Rey, \Lambda_d) = (\mathcal{M}_0\alpha, \Rey_0\alpha^{1/2}, \infty)$ in the limit $\alpha \rightarrow \infty$ for fixed $\mathcal{M}_0,\Rey_0$. In particular, we wish to evaluate $\eta_f^{\infty}:=\lim_{\alpha \rightarrow \infty}\eta_f(\mathcal{M}_0\alpha, \Rey_0\alpha^{1/2}, \infty)$. It is convenient here to write the similarity solution ODEs in the form obtained directly by substituting the self-similarity ansatz (\ref{eq:region_III_ansatz}) into the thin-film equations (\ref{eq:tf_flow_total}):
\begin{subeqnarray}
    -\frac{1}{2}U-\frac{1}{2}\eta\frac{dU}{d\eta}+U\frac{dU}{d\eta} &=&\frac{1}{2 \eta_f^4\mathcal{M}}\frac{d }{d \eta}\left(\frac{1}{\eta}\frac{d}{d \eta}\left(\eta\frac{d H}{d\eta}\right)\right)\nonumber\\
    &&+ \frac{4}{\Rey \eta_f^2}\frac{1}{H}\left(\frac{d}{d\eta}\left(\frac{H}{\eta}\frac{d}{d \eta}\left(\eta U\right)\right) - \frac{1}{2}\frac{U}{\eta}\frac{dH}{d\eta}\right),\slabel{eq:momentum_before_sub}\\
    -\frac{1}{2}\eta \frac{dH}{d\eta}&=&-\frac{UH}{\eta}-\frac{dU}{d\eta}H - U \frac{dH}{d\eta}, \slabel{eq:mass_before_sub}
    \label{eq:before_sub_high_M_Re_lim}
\end{subeqnarray} 
which is equivalent to the form given in the main text (\ref{eq:soluble_regIII}).

\begin{figure}
\begin{center}
\includegraphics[width=\textwidth]{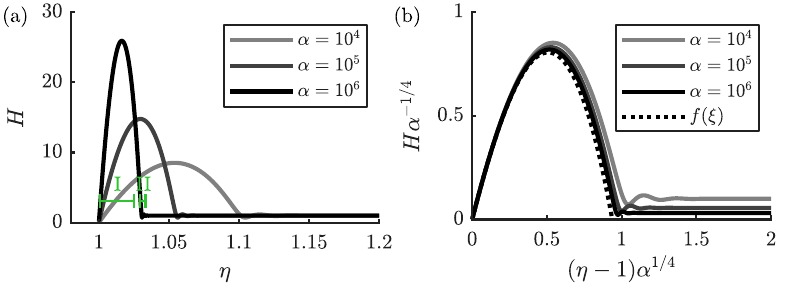}
\end{center}
\caption{Similarity solution profile in a high $\mathcal{M},\Rey$ limit. (a) Solid curves show the solution for $H$ to the similarity solution ODEs (\ref{eq:soluble_regIII}, \ref{eq:soluble_regIII_BC}) for $(\mathcal{M},\Rey,\Lambda_d) = (\mathcal{M}_0\alpha, \Rey_0\alpha^{1/2},\infty)$ where $\mathcal{M}_0=1,\Rey_0 = 10$ and $\alpha = 10^4,10^5,10^6$ (grey, dark grey, black). There are two distinct regions: region I which is a region of high curvature near $\eta = 1$ and region II connects region I to the portion of the solution for which $H = 1$. The vertical axis is given by $H$ and the horizontal axis is given by $\eta$ (b) Same as (a) but rescaled axes according to the scaling of region I. The dotted curve shows the analytical prediction (\ref{eq:f_high_lim}) in the limit $\alpha \rightarrow \infty$.}
\label{fig:simil_high_alpha}
\end{figure}

Upon looking at figure \ref{fig:simil_high_alpha}(a), it can be seen that there is a region of sharp curvature localised around $\eta = 1$, which we refer to as region I. Thus, we expect asymptotic and self-similar behaviour in $\alpha$ for $|\eta-1| \ll 1$. The scaling in region I for the velocity satisfies $U \sim \alpha^0$ from the boundary condition (\ref{eq:soluble_U_BC}). Then, balancing inertia and capillary balance in (\ref{eq:momentum_before_sub}) and global conservation of mass (\ref{eq:shooting_constraints}) gives $H \sim \alpha^{1/4}$ and $\Delta \eta \sim \alpha^{-1/4}$ where $\Delta \eta$ is the characteristic width of the $\eta$ coordinate. Then, the self-similarity ansatz for $H$ is given by
\begin{equation}
    H = \alpha^{\frac{1}{4}}f(\xi_{\text{I}}),~ \xi_{\text{I}}=\alpha^{\frac{1}{4}}(\eta-1).\label{eq:ansatz_reg_I_high_M_Re_limit}
\end{equation}
Then, (\ref{eq:momentum_before_sub}) gives to leading order in $\alpha$:
\begin{equation}
    -\frac{1}{4}=\frac{1}{2(\eta_f^{\infty})^4 \mathcal{M}_0}\frac{d^3f}{d\xi_{\text{I}}^3}\label{eq:cubic}
\end{equation}
There is a region which connects region I to the undisturbed sheet $H = 1, U = 0$, which we refer to as region II. Since $H = 1$ at far-field, in region II, it follows that $H \sim \alpha^0$. By matching the slope $dH/d\eta$ onto region I, where (\ref{eq:ansatz_reg_I_high_M_Re_limit}) gives that $dH/d\eta \sim \alpha^{1/2}$, it then follows that $\Delta \eta \sim \alpha^{-1/2}$. Then, conservation of momentum (\ref{eq:momentum_before_sub}) gives that $U \sim \alpha^0$. Then, letting the end of region I be given by $\xi_{\text{I}} =\xi_{\text{I}}^{\text{end}}$, the coordinate of region II is given by
\begin{equation}
    \xi_{\text{II}}:=\alpha^{\frac{1}{2}}\left(\eta-1-\xi_{\text{I}}^{\text{end}}\alpha^{-\frac{1}{4}}\right).
\end{equation}
Then, to leading order in $\alpha$, (\ref{eq:before_sub_high_M_Re_lim}) gives (upon rearrangement): 
\begin{subeqnarray}
    -\frac{1}{2}\frac{d}{d\xi_{\text{II}}}(UH)+\frac{d}{d\xi_{\text{II}}}(U^2H)&=&\frac{1}{2(\eta_f^{\infty})^4M_0}H\frac{d^3H}{d \xi_{\text{II}}^3} + \frac{4}{\Rey_0 (\eta_f^{\infty})^2}\frac{d}{d \xi_{\text{II}}}\left(H \frac{dU}{d\xi_{\text{II}}}\right).\slabel{eq:H_high_lim_eq}\\
    -\frac{1}{2}\frac{dH}{d\xi_{\text{II}}}+H\frac{dU}{d\xi_{\text{II}}}+U \frac{dH}{d\xi_{\text{II}}}&=&0.\slabel{eq:U_high_lim_eq}
\end{subeqnarray}
Integrating (\ref{eq:H_high_lim_eq}) and applying the boundary condition that $U = 0$ and $H = 1$ at far-field:
\begin{equation}
    U = \frac{1}{2}\left(1-H^{-1}\right).\label{eq:U_reg_II_high_M_Re_lim}
\end{equation}
Integrating (\ref{eq:U_high_lim_eq}), substituting (\ref{eq:U_reg_II_high_M_Re_lim}) and applying far-field boundary conditions gives
\begin{equation}
    -\frac{1}{4}+\frac{1}{4}H^{-1}=\frac{1}{2(\eta_f^{\infty})^4\mathcal{M}_0}\left(H\frac{d^2H}{d\xi_{\text{II}}^2}-\frac{1}{2}\left(\frac{dH}{d\xi_{\text{II}}}\right)^2\right) + \frac{2}{\Rey_0 (\eta_f^{\infty})^2}\left(H^{-1}\frac{dH}{d\xi_{\text{II}}}\right).
\end{equation}
In particular, it follows that
\begin{equation}
\lim_{\xi_{\text{II}}\rightarrow - \infty} \frac{1}{2(\eta_f^{\infty})^4\mathcal{M}_0}\left(H\frac{d^2H}{d\xi_{\text{II}}^2}-\frac{1}{2}\left(\frac{dH}{d\xi_{\text{II}}}\right)^2\right)=-\frac{1}{4}.
\end{equation}

We now have all the boundary conditions required to integrate region I (\ref{eq:cubic}). The boundary conditions are
\begin{equation}
    f(0)=0,~f'(0) = \sqrt{8\mathcal{M}_0},~ \int_0^{\xi_{\text{I}}^{\text{end}}}f d\xi_{\text{I}} = \frac{1}{2},~ f(\xi_{\text{I}}^{\text{end}})=0,~ \left(f'\left(\xi_{\text{I}}^{\text{end}}\right)\right)^2 = (\eta_f^{\infty})^4 \mathcal{M}_0,
\end{equation}
which are the five conditions required to solve exactly for the third-order ODE (\ref{eq:cubic}) with two unknowns $\eta_f^{\infty}$ and $\xi_{\text{I}}^{\text{end}}$.

\begin{figure}
\begin{center}
\includegraphics[width=\textwidth]{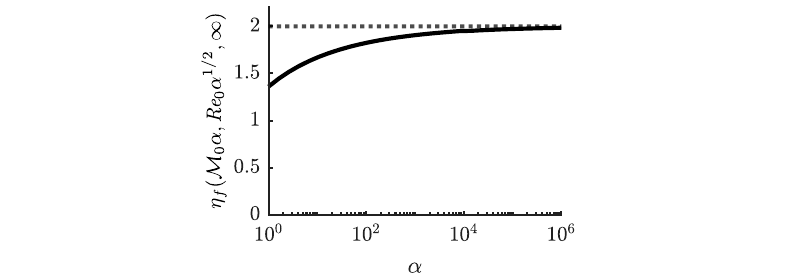}
\end{center}
\caption{Similarity solution prefactor in a high $\mathcal{M}, \Rey$ limit. Solid curve shows $\eta_f(\mathcal{M}_0\alpha, \Rey_0\alpha^{1/2},\infty)$ from numerical solutions to the similarity solution ODEs (\ref{eq:soluble_regIII}, \ref{eq:soluble_regIII_BC}) from $\alpha = 1$ to $\alpha = 10^6$ for $\mathcal{M}_0 = 1, \Rey_0 = 10$. Dotted line shows the analytical prediction in the limit $\alpha \rightarrow \infty$ where $\eta_f^{\infty}=2$ (\ref{eq:app_eta_f_lim}).}
\label{fig:eta_f_limit}
\end{figure}

After some algebra, it can be shown exactly that the solution is given by
\begin{equation}
    f = -\frac{4}{3}\mathcal{M}_0\xi^3 + \frac{(4-4\sqrt{2})\sqrt{2}}{\sqrt{3}(2-\sqrt{2})^{1/2}}\mathcal{M}_0^{3/4}\xi^2 + \sqrt{8\mathcal{M}_0}\xi.\label{eq:f_high_lim}
\end{equation}
with the constants evaluated to be
\begin{equation}
    \xi_{\text{I}}^{\text{end}} = \sqrt{3}\left(1-\frac{\sqrt{2}}{2}\right)^{\frac{1}{2}}\mathcal{M}_0^{-1/4}, ~\eta_f^{\infty}=2.\label{eq:app_eta_f_lim}
\end{equation}
There is good agreement between the numerical solution of the similarity ODEs (\ref{eq:soluble_regIII}, \ref{eq:soluble_regIII_BC}) and the analytical expression (\ref{eq:f_high_lim}) for $\alpha \gg 1$ (figure \ref{fig:simil_high_alpha}(b)). The prediction $\eta_f^{\infty}=2$ can also be seen by plotting $\eta_f(\mathcal{M}_0\alpha, \Rey_0\alpha^{1/2}, \infty)$ against $\alpha$ (figure \ref{fig:eta_f_limit}).

\section{Similarity solution for $\Lambda_d = 0$}\label{app:scaling_Ld0}

In this appendix, we show that the minimum thickness $h_{\text{min}} \sim \textit{Bi}^{-1}t^{-3}$ when $\Lambda_d = 0$.
We do not present the full similarity solution here: numerical difficulties in resolving the sharp surfactant front prevented us from fully verifying our analytical predictions against simulations. 
As for the infinitely soluble case \S \ref{subsec:inf_soluble}, a code specifically designed to account for strong surface tension gradients, such as that by \cite{Kitavtsev18} should allow for further investigation. We can however solve for the similarity solution in the region behind the surfactant front. 

The scalings for the region behind the surfactant front (i.e., region I) are as follows. First, the surface surfactant transport equation (\ref{eq:surface_surf_Ld0}) gives $\Gamma = c + \textit{O}(\textit{Bi}^{-1}t^{-1})$ at late times. Now, $\partial \Gamma/\partial r = 0$ to leading order from the momentum equations (\ref{eq:tf_mom}) and hence $\Gamma = c_0 := c(r=0,t)$ to leading order as $c = c_0$ at $r = 0$. Note that $c(r=0,t)$ is invariant over time (\ref{eq:bulk_surf_Ld0}). Thus, $c =c_0 + \textit{O}(\textit{Bi}^{-1}t^{-1})$. In other words, $c \approx c_0$ throughout the region behind the front and hence the radial Lagrangian coordinate $s$ (see Appendix \ref{app:horiz_Lag_coords}) satisfies $s\ll 1$ in the region since $c \approx c_0$ implies that the material elements near the origin initially form the region behind the surfactant front. Then, expanding in a Taylor series for $s \ll 1$ gives $c = c_0 + (\partial^2 c/\partial s^2)|_{s=0} (s^2/2)+\textit{O}(s^3)$, so that $s^2 = \textit{O}(\textit{Bi}^{-1}t^{-1})$. Thus, from (\ref{eq:horiz_lag_func}), $hr^2 = \textit{O}(\textit{Bi}^{-1}t^{-1})$. As for the infinitely soluble case \S \ref{subsec:inf_soluble}, balancing the Marangoni stress and advection terms in the momentum flux across the front gives $\partial \Gamma/\partial r \sim \partial (hu^2)/\partial r$. Then, the Rankine-Hugoniot relations implies that the surface surfactant concentration jump $c_0 \sim \dot{r}_f^2$ (with dots denoting time derivatives and $r_f$ the surfactant front), which gives $r_f = \textit{O}(t)$. Unraveling the scaling relations identified above, the characteristic scales of variables in the region behind the surfactant front are given by
\begin{equation}
    (r,~u,~h,~\Gamma-c_0,~c-c_0) \sim (t,~1,~\textit{Bi}^{-1}t^{-3}, ~\textit{Bi}^{-1}t^{-1}, ~\textit{Bi}^{-1}t^{-1}).
\end{equation}

With the scaling, we can solve for the similarity solution with the ansatz given by
\begin{subeqnarray}
    u &=& \eta_f V(\eta)\\ 
    h &=& \eta_f^{-2}\textit{Bi}^{-1}t^{-3} H(\eta)\\ 
    \Gamma &=& c_0 + \textit{Bi}^{-1} t^{-1} c_0G(\eta)\\
    c &=& c_0 + \textit{Bi}^{-1} t^{-1}c_0C(\eta)\label{eq:ansatz_Ld0}
\end{subeqnarray}
with the similarity coordinate
\begin{equation}
    \eta = \frac{r}{\eta_f t}.
\end{equation}
The prefactors of (\ref{eq:ansatz_Ld0}) are chosen for analytical convenience. Since we do not solve for the full similarity solution, we cannot solve for $\eta_f$, but we expect $\eta_f$ to be independent of $\textit{Bi}$, as discussed below.

With the ansatz (\ref{eq:ansatz_Ld0}), the thin-film equations (\ref{eq:tf_mom}, \ref{eq:tf_mass}, \ref{eq:surface_surf_Ld0}, \ref{eq:bulk_surf_Ld0}) become to leading order respectively
\begin{subeqnarray}
    \frac{dG}{d\eta}&=&0 \slabel{eq:mom_Ld0_simil}\\
    -3H - \eta \frac{dH}{d\eta}&=&-\frac{1}{\eta}\frac{d}{d\eta}\left(\eta V H\right)\slabel{eq:mass_Ld0_simil}\\
    0 &=& - \frac{dV}{d\eta}-\frac{V}{\eta}+C-G\slabel{eq:surface_surf_Ld0_simil}\\
    -C - \eta \frac{dC}{d\eta}+V\frac{dC}{d\eta} &=&0\slabel{eq:bulk_surf_Ld0_simil}
\end{subeqnarray}
Then, (\ref{eq:mom_Ld0_simil}) gives that $G = G_0$ for some constant $G_0$. Further, evaluating (\ref{eq:mass_Ld0_simil}, \ref{eq:surface_surf_Ld0_simil}) at $\eta = 0$ upon noting that $C(0)=0$ (as $c = c_0$ for all time at $r = 0$) and $H'(0)=0$ (axisymmetry), we have that
\begin{subeqnarray}
    -3H(0) = -H(0) \left.\left(V'+\frac{V}{\eta}\right)\right|_{\eta=0}\\
    0 = -\left.\left(V'+\frac{V}{\eta}\right)\right|_{\eta=0} -G_0
\end{subeqnarray}
and hence 
\begin{equation}
    G =G_0= -3.\label{eq:G_soln_Ld0}
\end{equation}
Then, eliminating $V$ from (\ref{eq:surface_surf_Ld0_simil}, \ref{eq:bulk_surf_Ld0_simil}), it follows that
\begin{equation}
    \eta \left(\frac{dC}{d\eta}\right)^2 = \frac{dC}{d\eta}-\eta \frac{d^2 C}{d\eta^2}
\end{equation}
which upon using $C(0)=0$ along with $C(1) = - \infty$ (due to the need to match onto surfactant front region) integrates to
\begin{equation}
    C = \log(1-\eta^2)\label{eq:C_soln_Ld0}
\end{equation}
and hence (\ref{eq:bulk_surf_Ld0_simil}) gives
\begin{equation}
    V = \eta - \frac{(1-\eta^2)\log(1-\eta^2)}{2\eta}.\label{eq:V_soln_Ld0}
\end{equation}

Finally, (\ref{eq:mass_Ld0_simil}) and (\ref{eq:bulk_surf_Ld0_simil}) can be rearranged to
\begin{equation}
    (V-\eta)\frac{dH}{d\eta}=-\left(-3 + \frac{dV}{d\eta}+\frac{V}{\eta}\right)H = -CH
\end{equation}
and hence from (\ref{eq:bulk_surf_Ld0_simil}),
\begin{equation}
    \frac{\frac{dH}{d\eta}}{H}=-\frac{dC}{d\eta}.
\end{equation}
Thus $H = H(0) e^{-C}$ and hence $H = H(0)(1-\eta^2)^{-1}$. Furthermore, $H(0)$ can be evaluated as follows. By change of coordinates from $(r,t)$ to radial Lagrangian coordinates $(s,t)$, it follows that
\begin{equation}
    \left.\frac{\partial^2 c}{\partial r^2}\right|_{(r=0,t)} = h(0,t) \left.\frac{\partial^2 c}{\partial s^2}\right|_{(s=0,t)}.\label{eq:eulerian_lagrangian_origin}
\end{equation}
Since (\ref{eq:bulk_surf_Ld0})  in Lagrangian coordinates gives $Dc/Dt=0$, substituting the ansatz (\ref{eq:ansatz_Ld0}) into (\ref{eq:eulerian_lagrangian_origin}) leads to
\begin{equation}
    \textit{Bi}^{-1}\eta_f^{-2}t^{-3}c_0 \left.\frac{d^2C}{d\eta^2}\right|_{\eta = 0} = \textit{Bi}^{-1}\eta_f^{-2}t^{-3}H(0) \left.\frac{\partial^2 c_i}{\partial s^2}\right|_{(s=0,t)}.
\end{equation}
The solution for $C$ (\ref{eq:C_soln_Ld0}) satisfies $(d^2C/d\eta^2)|_{\eta=0} = -2$ and hence
\begin{equation}
    H = \frac{2c_0}{\left(-\left.\frac{d^2 c_i}{d s^2}\right|_{s=0}\right)}\frac{1}{1-\eta^2}.\label{eq:H_soln_Ld0}
\end{equation}
Thus, for the initial condition of a Gaussian $c_i = e^{-r^2}$, it follows that $H(0) = 1$ and hence $h_{\text{min}} = \eta_f^{-2} \textit{Bi}^{-1}t^{-3}$. We do not consider the case where there is a local minimum at the origin $d^2 c_i/d s^2|_{s=0}>0$.

In summary, the similarity solution for the region behind the surfactant front is given by (\ref{eq:G_soln_Ld0}, \ref{eq:C_soln_Ld0}, \ref{eq:V_soln_Ld0}, \ref{eq:H_soln_Ld0}). The constant $\eta_f$ remains to be evaluated by matching the different spatial regions of the solution. Since the $\textit{Bi}$ parameter physically should not appear in other regions, we expect $\eta_f$ to be independent of $\textit{Bi}$ and hence $h_{\text{min}}\sim \textit{Bi}^{-1}t^{-3}$.

\begin{figure}
\begin{center}
\includegraphics[width=\textwidth]{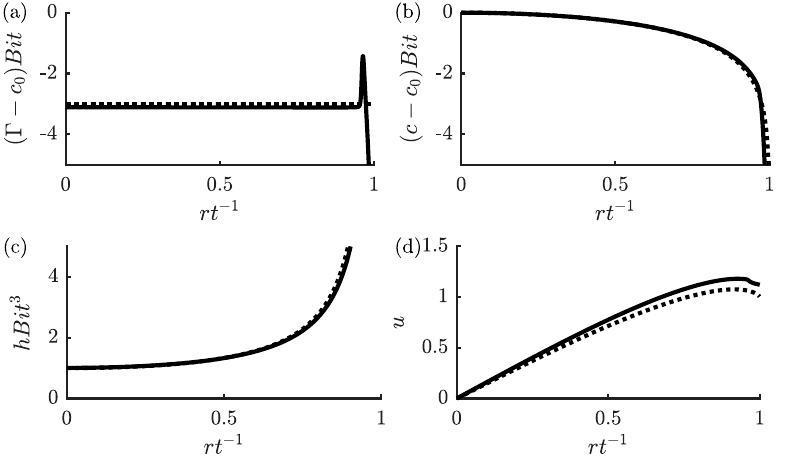}
\end{center}
\caption{Similarity solution profiles in the region behind the surfactant front for $\Lambda_d = 0$. Solid curves show the numerical solution to the thin-film equations (\ref{eq:tf_mom}, \ref{eq:tf_mass}, \ref{eq:surface_surf_Ld0}, \ref{eq:bulk_surf_Ld0}) for $\mathcal{M}=1,\Rey = 10, \textit{Bi}=10, \Lambda_d = 0$, $\Gamma_i = c_i = e^{-r^2}$ at $t = 5$. The similarity solution prediction (\ref{eq:G_soln_Ld0}, \ref{eq:C_soln_Ld0}, \ref{eq:V_soln_Ld0}, \ref{eq:H_soln_Ld0}) with the fit  $\eta_f = 1$ is shown by dotted curves. The plots are scaled according to the similarity solution (a) Interfacial surfactant concentration $\Gamma$ against the radial coordinate $r$. (b) Bulk surfactant concentration $c$ against $r$. (c) Thickness $h$ against $r$. (d) Velocity $u$ against $r$. Note that $c_0:=c(0,t)=c(0,0)$ is the value of $c$ at the origin. The results in all the panels are nondimensional (\ref{eq:nondim}). }
\label{fig:simil_Ld_0}
\end{figure}

The similarity solution profiles (\ref{eq:G_soln_Ld0}, \ref{eq:C_soln_Ld0}, \ref{eq:V_soln_Ld0}, \ref{eq:H_soln_Ld0}) are verified in figure \ref{fig:simil_Ld_0}. The solid curves show the numerical solution for $\textit{Bi}=10$ and $t = 5$. The dotted curves show the similarity solution predictions. We see that there is indeed good agreement upon taking $\eta_f = 1$ as a fitting parameter. 

\begin{figure}
\begin{center}
\includegraphics[width=\textwidth]{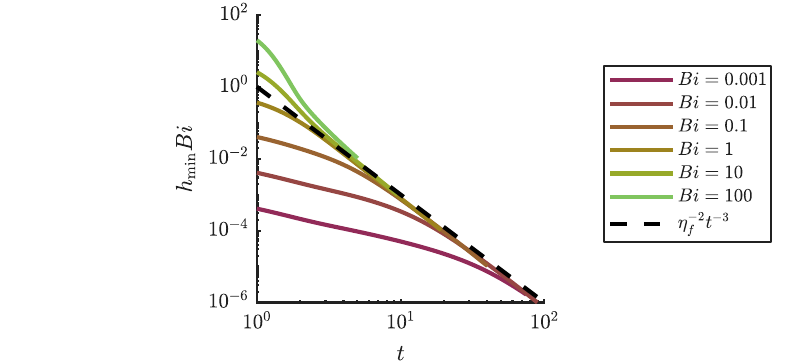}
\end{center}
\caption{Scaling of the minimum thickness $h_{\text{min}}\sim \textit{Bi}^{-1}t^{-3}$. Solid curves show numerical solutions to the thin-film equations (\ref{eq:tf_mom}, \ref{eq:tf_mass}, \ref{eq:surface_surf_Ld0}, \ref{eq:bulk_surf_Ld0}) for $\textit{Bi}=0.001,0.01,0.1,1,10,100$ with $\mathcal{M}=1,\Rey = 10, \Lambda_d = 0$, $\Gamma_i = c_i = e^{-r^2}$. The similarity solution prediction $h_{\text{min}}=\eta_f^{-2}t^{-3}$ is shown by a dashed line where $\eta_f = 1$ is a fit. The results in all the panels are nondimensional (\ref{eq:nondim}).}
\label{fig:simil_Ld_0_hmin}
\end{figure}

The scaling $h_{\text{min}}\sim \textit{Bi}^{-1}t^{-3}$ can be seen numerically by plotting $h_{\text{min}}\textit{Bi}$ against $t$, which is shown in figure \ref{fig:simil_Ld_0_hmin}. The coloured curves show numerical solutions for $\textit{Bi}=100, \cdots 0.001$ (the same solution curves as in figure \ref{fig:regime_Ld0}). The dashed line shows the analytical prediction $h_{\text{min}}=\eta_f^{-2}t^{-3}$ where the value of $\eta_f$ is once again fitted to $\eta_f = 1$. 

\bibliographystyle{jfm}

\bibliography{mybib.bib}
\end{document}